\documentclass[fleqn,usenatbib]{mnras}

\usepackage{newtxtext,newtxmath}

\usepackage[T1]{fontenc}

\DeclareRobustCommand{\VAN}[3]{#2}
\let\VANthebibliography\thebibliography
\def\thebibliography{\DeclareRobustCommand{\VAN}[3]{##3}\VANthebibliography}

\usepackage{graphicx}	% Including figure files
\usepackage{amsmath}	% Advanced maths commands

\newcommand{\kmps}{\rm km~s\ensuremath{^{-1} }\,}

\newcommand{\Msun}{M\ensuremath{_\odot}}

\newcommand{\FeH}{\rm [Fe/H]}

\newcommand{\mymodels}[1]{%
    \IfEqCase{#1}{%
        {40}{{\tt S2}}%
        {41}{{\tt S1}}%
        {42}{{\tt M1}}%
        {43}{{\tt M2}}%
        {44}{{\tt ISO}}%
        {45}{{\tt M1noG}}%
        {46}{{\tt M2noG}}%
        {47}{{\tt S1noG}}%
        {48}{{\tt S2noG}}%
    }[\PackageError{tree}{Undefined option to tree: #1}{}]%
}%

\title[Impact of orbiting satellites on star formation]{Impact of orbiting satellites on star formation rate evolution \\ and metallicity variations in Milky Way-like discs}

\author[B. Annem \& S. Khoperskov]{
Bhargav Annem$^{1}$\thanks{E-mail: bhargav.annem@gmail.com}
and Sergey Khoperskov$^{2,3}$ \\
$^{1}$ California Institute of Technology, 1200 E California Blvd, Pasadena, CA 91125, USA  \\
$^{2}$ Leibniz Institut f\"{u}r Astrophysik Potsdam (AIP), An der Sternwarte 16, D-14482, Potsdam, Germany\\
$^{3}$ GEPI, Observatoire de Paris, Université PSL, CNRS, 5 Place Jules Janssen, 92190 Meudon, France
}

\pubyear{2022}

\begin{document}
\label{firstpage}
\pagerange{\pageref{firstpage}--\pageref{lastpage}}
\maketitle

\begin{abstract}
%The Milky Way galaxy is surrounded by dozens of satellite galaxies whose close passages perturb the disc externally. 
At least one major merger is currently taking place in the Milky Way~(MW). The Sgr dwarf spheroidal galaxy is being tidally destroyed while orbiting around the MW, whose close passages perturb the MW disc externally. In this work, using a series of hydrodynamical simulations, we investigate how massive dwarf galaxies on quasi-polar Sgr-like orbits impact the star formation activity inside the MW-like discs. First, we confirm that interactions with orbiting satellites enhance the star formation rate in the host galaxy. However, prominent short-time scale bursts are detected during the very close passages~($<20$~kpc) of massive~($>2\times 10^{10}$~\Msun) gas-poor satellites. In the case of gas-rich satellites, while we see a substantial enhancement of the star formation on a longer time scale, we do not detect prominent peaks in the star formation history of the host. This can be explained by the steady accretion of gas being stripped from the satellite, which smoothens short-term variations in the star formation rate of the host. It is important that the impact of the satellite perturbations, especially its first encounters, is seen mainly in the outer~($>10$~kpc) disc of the host. We also found that the close passages of satellites cause the formation of a substantial amount of low-metallicity stars in the host, and the effect is the most prominent in the case of gas infall from the satellites resulting in the dilution of the mean stellar metallicity soon after the first pericentric passage of massive satellites. Our simulations are in favour of causality between the recent passages of the Sgr galaxy and the bursts of the star formation~(SF) in the solar neighbourhood~($\approx 1 $ and $\approx 2$~Gyr ago); however, in order to reproduce the SF burst at its first infall~($\approx 6$~Gyr), we require a very close pericentric passage~($<20$~kpc) with subsequent substantial mass loss of the Sgr precursor. 
\end{abstract}


%%%%%%%%%%%%%%%%%%%%%%%%%%%%%%%%%%%%%%%%%%%%%%%%%%%%%%%%%%%%%%%
\begin{figure*}
\begin{center}
\includegraphics[width=1\hsize]{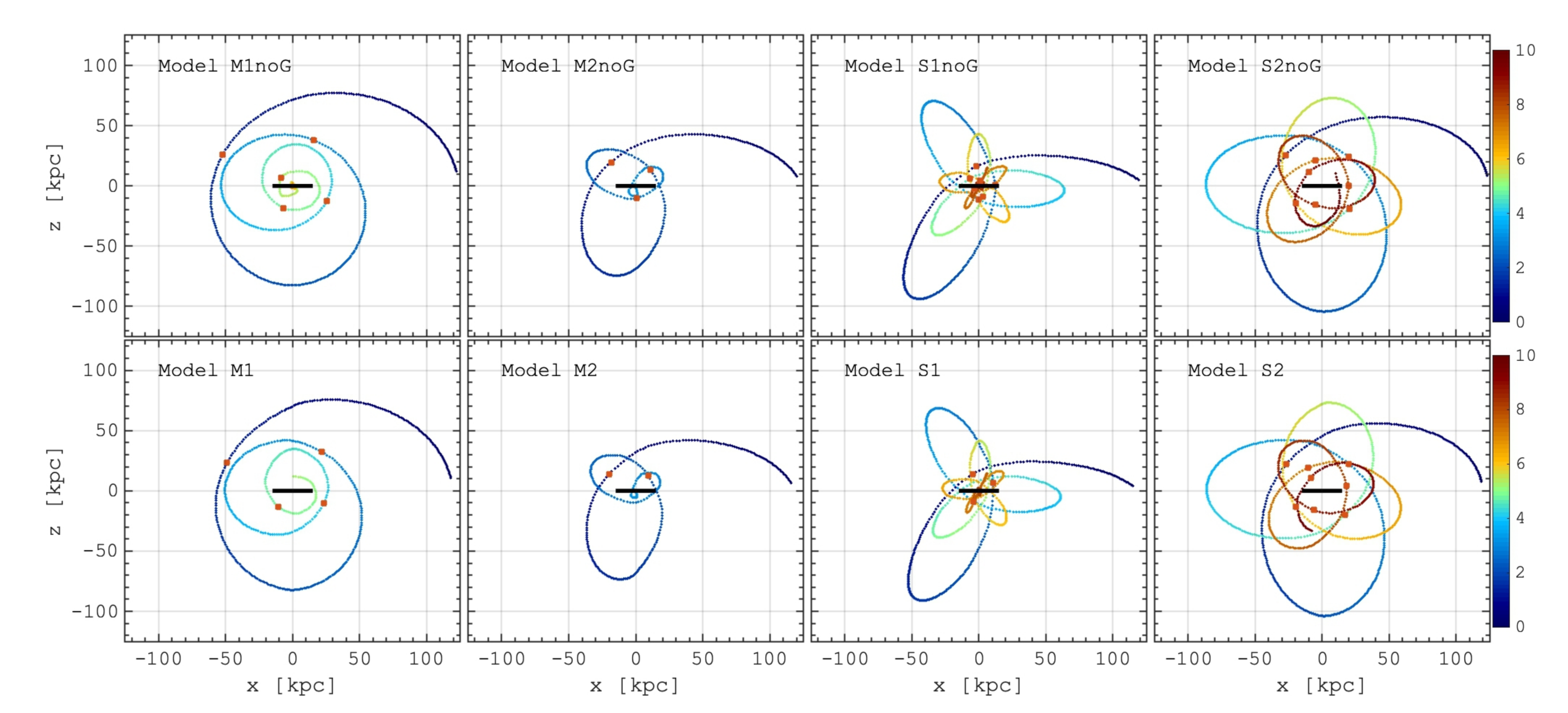}
\caption{Orbits of satellites in different models. The top row corresponds to the models with a gasless satellite, while the bottom shows the gas-rich satellites. The XZ-projections of the orbits are shown where the disc of the host galaxy is located in XY-plane, highlighted by the black line, representing a $15$~kpc radial scale. The orbits of the satellites are colour-coded by the integration time in Gyr as shown in the colour bars. The red squares highlight moments of pericentric passages of the orbiting satellites. }\label{fig::orbits}
\end{center}
\end{figure*}
%%%%%%%%%%%%%%%%%%%%%%%%%%%%%%%%%%%%%%%%%%%%%%%%%%%%%%%%%%%%%%%

\section{Introduction}
The Milky Way~(MW) is believed to be a typical disc galaxy with a central bar and boxy/peanut bulge in the center~\citep{1995ApJ...445..716D,2013MNRAS.435.1874W,2016AJ....152...14N} and extended, likely flocculent, multi-arm spiral structure~\citep{2005AJ....130..569V,2009ApJ...700..137R,2016ARA&A..54..529B}. The outer disc of the MW is warped relative to the disc midplane inside the solar circle~\citep{1957AJ.....62...93K,2001ApJ...556..181D,2018MNRAS.481L..21P}. The presence of the warp is believed to be a manifestation of the interaction with a massive galaxy-companion~\citep[see, e.g.][]{1998A&A...336..130I, 2003ApJ...583L..79B,2016ApJ...823....4D,2018MNRAS.473.1218L,2018MNRAS.481..286L}. Different models have been proposed to identify the mechanism of its formation, and it is generally believed that the Sagittarius dwarf galaxy~(Sgr)~\citep{1994Natur.370..194I,1995MNRAS.277..781I} could have been the most prominent external perturber of the MW disc~\citep{2003ApJ...583L..79B,2011MNRAS.414L...1M,2011Natur.477..301P,2013MNRAS.429..159G,2018MNRAS.478.3809S}. Although, some alternative ideas have not been ruled out~\citep{2019MNRAS.484.1050D,2019A&A...622L...6K}, it is supposed that the recent passages of the Sgr dwarf galaxy are responsible for the emergence of the phase-space snail in the solar neigbourhood~(SNd)~\citep[][but an extensive discussion of lacunae in this theory can be found in \cite{2021MNRAS.503..376B,2022ApJ...927..131B}]{2018Natur.561..360A,2019MNRAS.485.3134L,2019MNRAS.486.1167B}.

Recently, it was also proposed that recurrent perturbations of the MW disc caused by the Sgr dwarf galaxy can temporally enhance the star formation in the disc~\citep{2020NatAs...4..965R}. Using the Gaia DR2 data \citep{2018A&A...616A..10G}, \cite{2020NatAs...4..965R} recovered the star formation history~(SFH) in a local volume around the SNd. A striking feature of the recovered SFH~(or age distribution) is multiple distinct bursts~(at $1, 2$, and $\approx6$~Gyr ago) of the star formation~\citep[see, also][who found the SF bursts in the MW at $\approx2$ and $4-6$~Gyr ago]{2006A&A...459..783C}, which correlate with the predicted pericentric passages of the Sgr dwarf galaxy~\citep{2010ApJ...714..229L,2011Natur.477..301P,2015MNRAS.454..933D,2018MNRAS.481..286L}. It is important to note that there are some other models which expect the pericentric passages to happen at slightly different times~\citep[see, e.g.][]{2017MNRAS.464..794G,2017ApJ...836...92D,2018MNRAS.478.5263T,2022MNRAS.516.1685D}. The discrepancies between different models of the Sgr orbital motion naturally arise from the uncertainties in the mass-loss over time~\citep[see, e.g.][]{2000MNRAS.314..468J,2017MNRAS.464..794G,2021MNRAS.504.3168B}, thus affecting the dynamical friction strength and expected perturbations of the MW potential~\citep{2022MNRAS.512..739D}. This affects the Sgr stream used to constrain the orbit of the Sgr precursor~\citep{2015ApJ...802..128G,2021MNRAS.501.2279V}. Another approach to date the pericentric passages of dwarfs is based on the timing of star formation episodes inside these satellite galaxies~\citep[see, e.g.][]{2007ApJ...663..960S, 2011A&A...525A..99P, 2021MNRAS.502..642R}.  Indeed grasping the underlying physics driving increased star formation due to interactions holds profound significance, as it offers means to extract orbital parameters of interacting systems \citep[the SMC-LMC system][]{2022MNRAS.513L..40M} and it enables deeper insights into past events in the evolution of dwarf galaxies~\citep[e.g. Leo I][]{2021MNRAS.501.3962R}. In the case of the Sgr dwarf and its stream, different authors found several episodes of SF: $0.5$, $5$ and $11$~\citep{2000AJ....119.1760L}; $5-7$, $11$ and $13$~Gyr ago~\citep{2015MNRAS.451.3489D}; $2.3$, $4$ and $6$~Gyr ago~\citep{2007ApJ...667L..57S}. To summarize, several recent works report that both MW and Sgr precursor have experienced several bursts of SF in the last $\approx 6$~Gyr and some of these bursts could have happened at the moments of pericentric passages of the Sgr.

Generally speaking, different models suggest that interactions and mergers with sufficiently massive dwarf galaxies, under certain conditions, may enhance the star formation in the main galaxy~\citep{1991ApJ...370L..65B,1992ApJ...387..152J,1994ApJ...425L..13M,2005MNRAS.361..776S,2008MNRAS.384..386C,2014MNRAS.442L..33R,2015MNRAS.448.1107M,2016MNRAS.462.2418S,2022MNRAS.516.4922R}. This effect depends on the parameters of interaction such as the perturber's orbit, relative mass, and relative gas fraction~\citep{2007A&A...468...61D,2008A&A...492...31D,2013MNRAS.430.1901H,2015MNRAS.448.1107M,2017MNRAS.468.4189P}. For instance, \cite{2021MNRAS.506..531D} found a reasonable correlation between the peaks of SF and pericentric passages of massive dwarfs in the CLUES simulations. A similar effect was demonstrated using the HESTIA simulations of the Local Group~\citep{2022arXiv220604521K}, where it was suggested that not all close passages or flybys of dwarf galaxies enhance the SF~\citep[see also][]{2007A&A...468...61D,2009ApJ...706...67R,2014MNRAS.442L..33R}. Despite the mounting growth of arguments in favour of Sgr-driven bursts of SF in the MW, it is inherently difficult to disentangle the impact of multiple perturbers moving on different orbits from internal processes~(bar formation and evolution, spirals arms formation) in cosmological models. %, which makes the results rather suggestive relying on a correlation~(between close satellite passages and SF bursts) which may not necessarily mean causality. 

In this work, we aim to investigate the evolution of the star formation rate in controlled simulations of interaction between a MW-type disc galaxy and a massive dwarf galaxy on a nearly-polar orbit. Although, our work is inspired by the recent results suggesting a possible impact of the pericentric passages of the Sgr dwarf galaxy on the SF activity in the MW, we do not aim to reproduce the exact parameters of the Sgr dwarf galaxy and the observed configuration of its tidal streams.
The paper is organized as follows. In Section.~\ref{sec::model} we describe the initial conditions of our models and the simulations setup. In Sections~\ref{sec::results_sfr_nogas}-~\ref{sec::results_isfr} we analyze the SFHs features of the MW-like galaxies depending on the gas content, masses and pericentric distances of orbiting satellites. In Section.~\ref{sec::results_metallicity} we present the metallicity behaviour of stars in the host galaxy under influence of massive satellite. In Section~\ref{sec::discussion} we discuss the MW-Sgr interaction in the context of our simulations. Finally, in Section~\ref{sec::summary} we summarize our main findings.

%%%%%%%%%%%%%%%%%%%%%%%%%%%%%%%%%%%%%%%%%%%%%%%%%%%%%%%%%%%%%%%
\begin{figure*}
\begin{center}
\includegraphics[width=1\hsize]{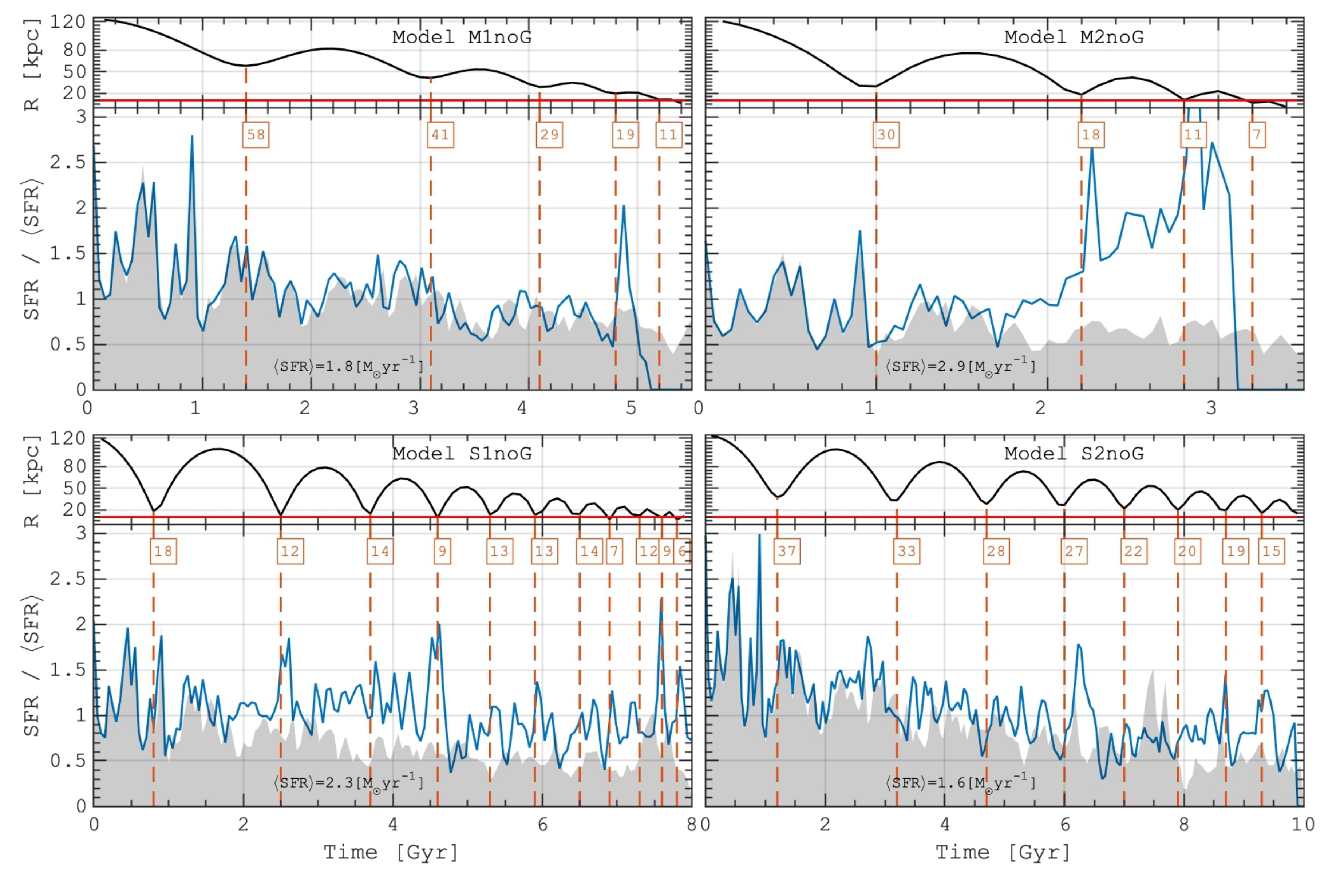}
\caption{Orbital decay of satellites and star formation histories of the host galaxies. The top sub-panels show the orbital decay of satellite galaxies, while the bottom ones show the evolution of the star formation history in the host galaxies. The SFHs are shown by the blue lines and compared to the SFH in the isolated host simulation, shown by the grey-filled area. For a better comparison between different models, in each panel, both SFHs are normalized by the mean SFR in the interacting case $\rm \langle SFR\rangle$, marked in each panel. The red vertical lines highlight moments of pericentric passages where the pericentric distance (in kpc) is shown in the corresponding red boxes.}\label{fig::sfr_nogas}
\end{center}
\end{figure*}
%%%%%%%%%%%%%%%%%%%%%%%%%%%%%%%%%%%%%%%%%%%%%%%%%%%%%%%%%%%%%%%
%%%%%%%%%%%%%%%%%%%%%%%%%%%%%%%%%%%%%%%%%%%%%%%%%%%%%%%%%%%%%%%
\begin{figure*}
\begin{center}
\includegraphics[width=1\hsize]{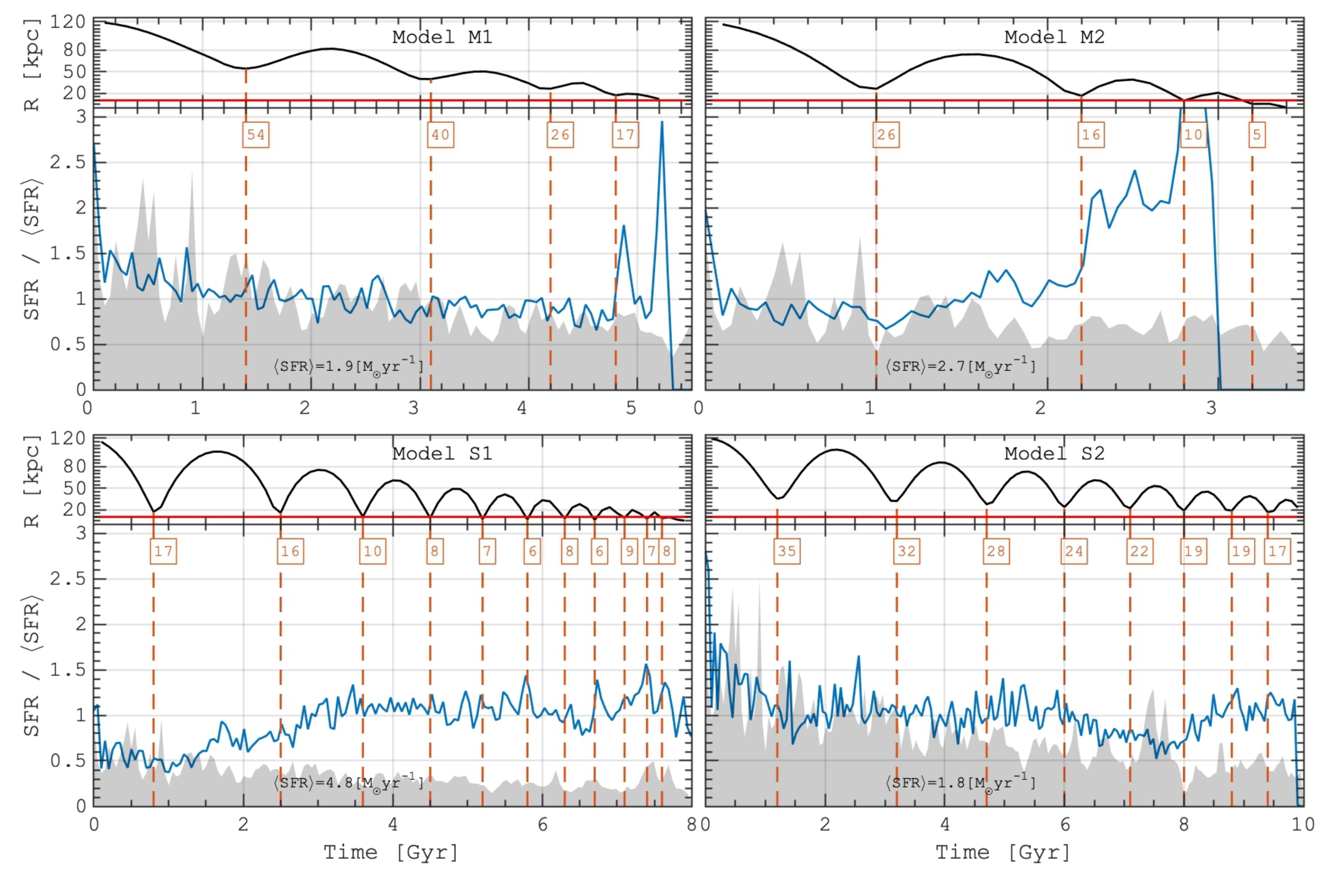}
\caption{Same as in Fig.~\ref{fig::sfr_nogas} but for gas-rich satellites. Note that the orbits gas-rich satellites are identical to the gasless satellites simulations.}\label{fig::sfr_gas}
\end{center}
\end{figure*}
%%%%%%%%%%%%%%%%%%%%%%%%%%%%%%%%%%%%%%%%%%%%%%%%%%%%%%%%%%%%%%%

%%%%%%%%%%%%%%%%%%%%%%%%%%%%%%%%%%%%%%%%%%%%%%%%%%%%%%%%%%%%%%%
\begin{figure*}
\begin{center}
\includegraphics[width=1\hsize]{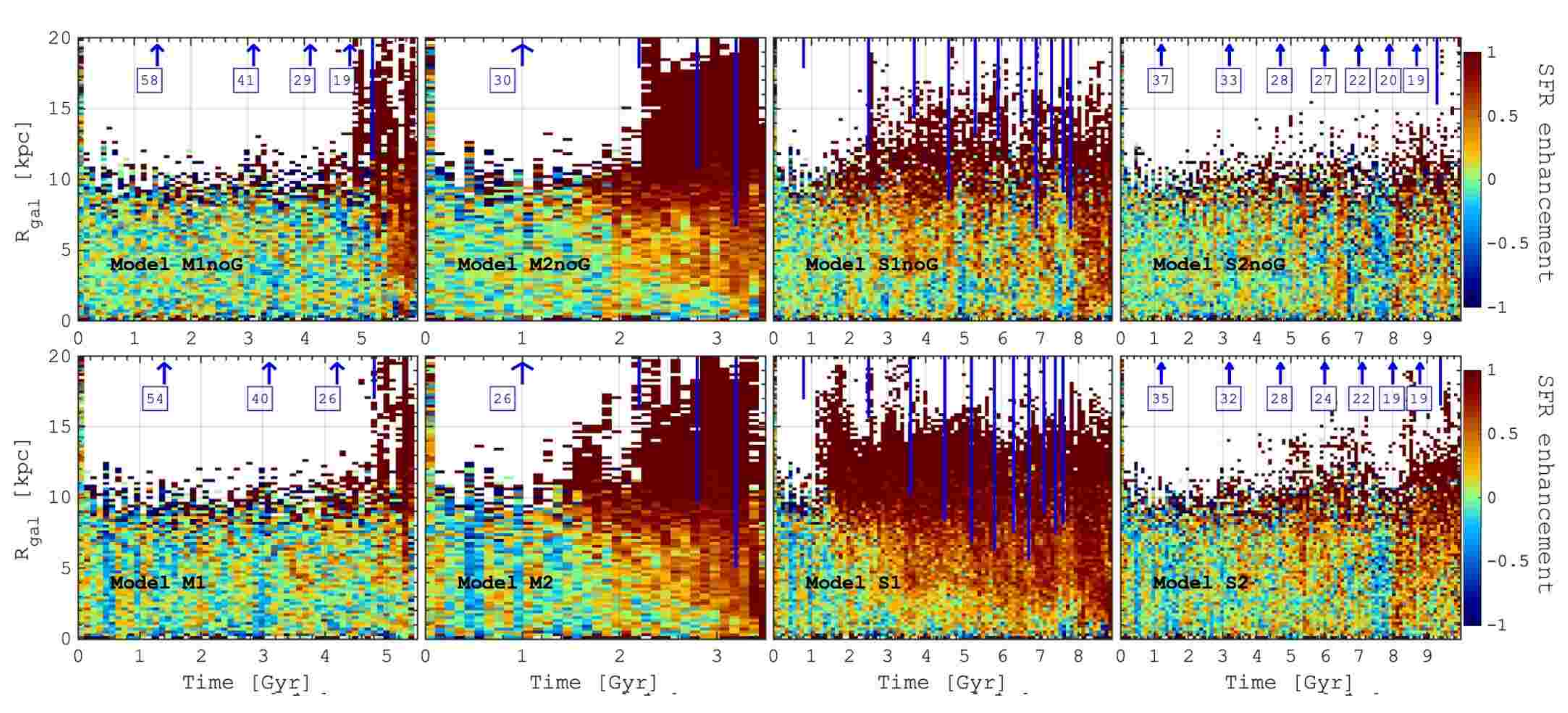}
\caption{Relative enhancement of the star formation in interacting models compared to the isolated one. The maps are colour-coded with following parameter $\rm \delta(R_{gal},t) = (\Sigma_{int}-\Sigma_{int})/(\Sigma_{int} + \Sigma_{iso})$ showing how much the star formation rate in the host galaxy is compared to the isolated model. The blue vertical lines highlight the moments of pericentric passages if the values are below $19$~kpc; otherwise, the pericenters are marked by the arrows with the pericentric distance in kpc. Note that the vertical axis shows the galactocentric distance $R_{gal}$ in cylindrical coordinates while the pericenters correspond to a 3D distance to the host centre.}\label{fig::relative_sfr}
\end{center}
\end{figure*}
%%%%%%%%%%%%%%%%%%%%%%%%%%%%%%%%%%%%%%%%%%%%%%%%%%%%%%%%%%%%%%%

\section{Models}\label{sec::model}

\subsection{Initial conditions for the host and dwarf galaxies}\label{sec::modelICs}
Initial conditions for the host and satellite galaxy were generated using the AGAMA software~\citep{2019MNRAS.482.1525V}. In all of our simulations, the initial setup for the host galaxy is identical and has three components~(dark matter, pre-existing stars, and gas), and has the following initial settings. The initial stellar disc is represented by a Miyamoto-Nagai density profile~\citep{1975PASJ...27..533M} with a characteristic scale length of $4$~kpc, vertical thicknesses of $0.3$~kpc, and a mass of $6 \times 10^{10}\Msun$. The simulation includes a live dark matter halo~($2\times 10^6$ particles) whose density distribution follows a Plummer sphere~\citep{1911MNRAS..71..460P} with a total mass of $9.2 \times 10^{11}~\Msun$ and a scale radius of $21$~kpc. The gas component initially features $10^6$ particles with a mass of $1 \times 10^4$~\Msun, which can be partially converted into new-formed stars. The gas is distributed in a thin disc component following a Miyamoto-Nagai density distribution with a scale length of $5$~kpc. This choice of parameters leads to a galaxy mass model with a circular velocity of $\approx 230$~\kmps. 

We study the impact of both high-mass and low-mass satellites with a total mass of $2\times 10^{10}~\Msun$ and $6 \times 10^{10}~\Msun$, respectively. The DM component of the satellite is represented by $2.5 \times 10^5$ particles redistributed following an NFW model with a scale length of $10$~kpc. Stars~($10^5$ particles) are initially distributed using a Plummer model with a scale length of $4$~kpc. The gas is considered as a Miyamoto-Nagai disc with a characteristic scale length of $2$~kpc with a mass of gas particles of $1 \times 10^4$~\Msun. The masses of the DM, stellar, and gas components are shown in Table.~\ref{tab::models}. In the table, we also provide the initial velocity of the satellite, which is located in the host galaxy plane at $x=125$~kpc from its center~\citep[similar to ][]{2017ApJ...836...92D,2018MNRAS.478.5263T}. In this work, we study the impact of the satellite on a polar orbit, which is similar to the orbit of the Sgr dwarf galaxy~\citep[e.g.][]{2015ApJ...802..128G,2017MNRAS.464..794G}. In total, we run a few dozens of different models where we varied the initial velocity component of the satellite in a wide range, but in this paper we only present $8$ models, which provide us with an orbit that is quantitatively similar to the Sgr and, especially in the case of massive satellites, do not rapidly merge. The full set of models considered here includes models with gas in a satellite where we can capture the effects of the gas accretion from the satellite together with gasless models~(inside the satellite) where we can study the purely kinematic impact on the host galaxy star formation. The star formation histories of the host galaxies are also compared to those of the isolated host galaxy evolution~(model \mymodels{44}).

In order to investigate the impact of the satellite passages on the metallicity distribution in the host galaxies, we initialize the radial metallicity profile of the interstellar medium~(ISM) in both host and satellite galaxies. Since newborn stars inherit their abundances from the gas component, radial redistribution and accretion of gas from the satellite would affect the metallicity distribution of the host. Since we do not aim to fit any particular data set in this work we adopted the negative metallicity gradients of $-0.13$ and $\rm -0.05\ dex/kpc$ with the central value of $0.1$ and $\rm -0.5\ dex$ for the host and satellite galaxy, respectively.  Nevertheless, the metallicity gradient for the host galaxy chosen is in agreement with some recent estimations for the MW disc at $\approx 10$ Gyr ago~\citep{2022arXiv221204515L,2023MNRAS.tmp.1561R}. The adopted metallicity gradient value for the dwarf galaxy is typical for nearby dwarf galaxies~\citep[see, e.g.][]{2022A&A...665A..92T}.

In this work, we run a series of simulations of galaxy mergers that used the hybrid $N$-body/hydro code GIZMO~\citep{2015MNRAS.450...53H,2017arXiv171201294H}. We invoked the meshless finite mass~(MFM) hydrodynamical solver and an adaptive gravitational softening for the gas. The \cite{2003MNRAS.339..312S} multi-phase interstellar matter~(ISM) algorithm has been used, and the feedback includes the TypeI and TypeII supernova. Following \cite{2013MNRAS.432.2647H,2017arXiv171201294H}, the star formation was allowed only in virialized regions.

%%%%%%%%%%%%%%%%%%%%%%%%%%%%%%%%%%%%%%%%%%%%%%%%%%%%%%%%%%%%%%%
\begin{table}
    \begin{tabular}{ |c|c|c| c| c| c|c| } 
    \hline
    Model  & Satellite & Satellite & Satellite & $v_x$  & $v_z$  \\ 
    name   & DM mass & stellar mass& gas mass &   \\ 
             & $10^{10}$ \Msun & $10^{10}$ \Msun  & $10^{10}$ \Msun & \kmps  & \kmps  \\ 
    \hline
    \mymodels{44} & - & - & - & - & - &  \\ 
    \mymodels{45} & 5 & 1 & - & -10 & 120  \\ 
    \mymodels{46} & 5 & 1 & - &  -30 & 80  \\ 
    \mymodels{47} & 1.6 & 0.4  & - &  -40 & 50  \\ 
    \mymodels{48} & 1.6 & 0.4 & - & 0 & 90  \\ 
    \mymodels{42} & 5 & 0.5 & 0.5 & -10 & 120  \\ 
    \mymodels{43} & 5 & 0.5 & 0.5 & -30 & 80  \\ 
    \mymodels{41} & 1.6 & 0.2 & 0.2 & -40 & 50  \\ 
    \mymodels{40} & 1.6 & 0.2 & 0.2 & 0 & 90 \\ 
    \hline
    \end{tabular}
    \caption{Parameters of models. The model {\tt ISO} is an isolated MW-type galaxy simulation while others include the dynamics of satellite galaxy of different mass and different gas amount. Initial coordinates of satellite galaxies are $[125,0,0]$ and $v_y=0$~\kmps, while the centre of the host galaxy is located at $[0,0,0]$. The parameters of the host galaxy are the same across the models and described in Sec.~\ref{sec::modelICs}.}
    \label{tab::models}
\end{table}
%%%%%%%%%%%%%%%%%%%%%%%%%%%%%%%%%%%%%%%%%%%%%%%%%%%%%%%%%%%%%%%

\subsection{Orbits of satellites}\label{sec::model_orbits}
In Fig.~\ref{fig::orbits}, we present the XZ-projections of the satellites' orbits where the host galaxy is placed in the XY plane. Remember that in this work, we study the interactions of the MW-type galaxies with satellites on a polar orbit. The top row corresponds to the models without gas in the satellites, and the bottom row shows the orbits of gas-rich satellites. Note that the orbits in these two cases are nearly identical because the total mass of the satellites is the same in both setups. Thus, they experienced a very similar evolution due to a comparable impact of dynamical friction dragging satellite galaxies toward the centre of the host. The first two columns show the orbits of massive satellites, while low-mass satellites are shown in the two rightmost columns~(see Table~\ref{tab::models} for the model parameters). In the case of massive satellites, the merger is completed on a relatively short time scale, and we do not study the evolution of these models after that $\approx 3-5$~Gyr. In this case, we have access to $3-4$ pericentric passages, marked with red squares in Fig.~\ref{fig::orbits}. Meanwhile, the low-mass satellites experience a long-term evolution, slowly swinging towards the centre of the host for about $8-10$~Gyr. In these models, we detected more than $8$ pericentric passages. In all of the models, the pericentric distances decrease with time, thus making us possible to study the impact of the satellites on the SF in the host galaxy at different distances in just a few models. Also note that although we do not aim to reproduce the motion of the Sgr dwarf galaxy, the orbits of our satellite galaxies are quite similar to some previous models of the Sgr orbit~\citep[see, e.g.][for reference]{2011Natur.477..301P,2017ApJ...836...92D,2018MNRAS.478.5263T}, thus suggesting that our conclusions can be extrapolated on the impact of the Sgr on the SF in the MW.

\section{Results}\label{sec::results}
In this section, we investigate the connection between the SF evolution of the MW-like hosts and recurrent perturbations by satellite galaxies orbiting around. 

\subsection{Models with no gas inside satellite galaxies}\label{sec::results_sfr_nogas}
First, we discuss the evolution of models with no gas inside satellite galaxies~(models \mymodels{45}, \mymodels{46}, \mymodels{47} and \mymodels{48}). In Fig.~\ref{fig::sfr_nogas}, we show the orbital decay of the satellite galaxies in different models together with the SFHs of the host galaxies. The SF in the host considers only stars that are being formed in the disc within $R_{gal}=25$~kpc and not beyond $2$~kpc from the midplane. Also, to ensure that we do not include any stars formed in the satellite galaxies inside this region, we count only stars if their absolute value of the initial vertical velocity is below $50$~\kmps. However, the latest criterion starts to be important at the very last episodes of the merger. In each panel, the star formation rate of the host galaxy~(blue line) is compared to the one in the isolated galaxy simulation~(grey area), where, in order to put all the panels on the same scale, we normalize both~(interacting and isolated) SFHs by the mean star formation rate~($\langle SFR \rangle$, see the values in each panel) in the interacting simulation. For each SFH plot, we show the red vertical lines that correspond to the pericentric passages of the dwarf galaxies, where the pericentric distances are marked accordingly.

%%%%%%%%%%%%%%%%%%%%%%%%%%%%%%%%%%%%%%%%%%%%%%%%%%%%%%%%%%%%%%%
\begin{figure*}
\begin{center}
\includegraphics[width=1\hsize]{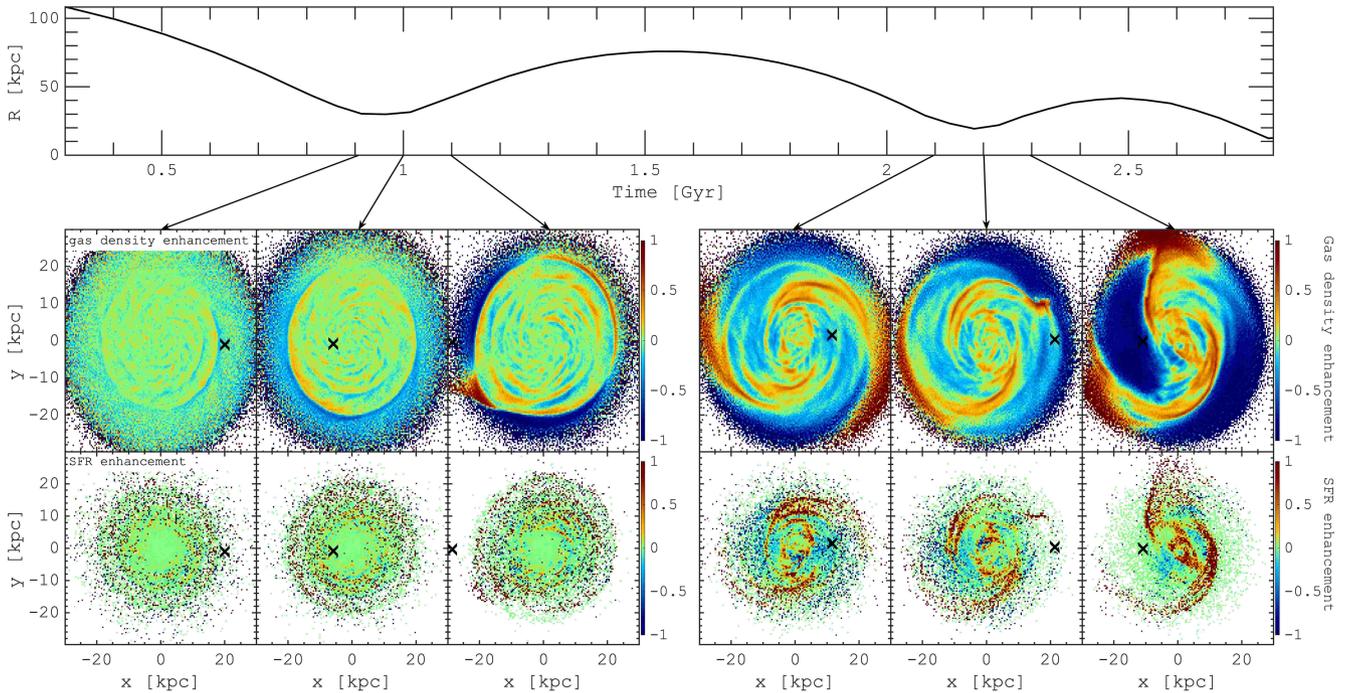}
\caption{Relative gas density and star formation enhancement in interacting models compared to the isolated one. {\it Top panel} shows the orbit of the satellite galaxy in model \protect\mymodels{46} with two pericentric passages before coalescence. {\it In bottom}, the maps are colour-coded as in Fig.~\ref{fig::relative_sfr}, showing how the gas density or star formation rate in the host galaxy is compared to the isolated model at the same times. The snapshots are chosen just before, during, and after a given pericentric passage.}\label{fig::relative_sfr_maps}
\end{center}
\end{figure*}
%%%%%%%%%%%%%%%%%%%%%%%%%%%%%%%%%%%%%%%%%%%%%%%%%%%%%%%%%%%%%%%

We start our analysis from the massive-dwarf simulations~($\approx 6\times 10^{10}~\Msun$) shown in the top row of Fig.~\ref{fig::sfr_nogas}. In model \mymodels{45}, we can see that the SFH is nearly identical to the isolated one until the latest pericentric passage~($19$~kpc) at $\approx 4.8$~Gyr. In this case, the SFR rapidly increased by a factor of $\approx 2$ compared to the isolated simulation. This suggests that even a very massive dwarf galaxy does not significantly affect the SFR in the host galaxy while it is orbiting at distances $>20$~kpc. The same result is seen in model \mymodels{46}, where the first passage, at $30$~kpc, has no impact on the SFR of the host. However, the next pericenter~($\approx 18$~kpc) corresponds to a prominent peak of the star formation. Note, however, that in this model, the star formation rate starts to increase gradually sometime before the second pericenter. The next sharp peak in the SFH also correlates with the next, even closer pericentric passage~($\approx 11$~kpc). Even with no extra gas infall from the satellite, the overall star formation rate is much higher compared to the isolated simulation, which suggests that only the gas of the host is involved in the star formation. Therefore, the increase of the star formation rate is due to the contraction of the gas caused by rapidly strengthened tidal forces~\citep{2009ApJ...706...67R,2014MNRAS.442L..33R,2022MNRAS.516.4922R} during the close passages of the satellite galaxy.

In the lower-mass satellite simulations~($\approx 1.2\times 10^{10}~\Msun$, see bottom row of Fig~\ref{fig::sfr_nogas}), we can see a similar behaviour of the star formation as above. For instance, in model \mymodels{47}, we see many peaks of the SFR corresponding to a number of the pericentric passages, except for the first one at $18$~kpc. This simply shows that lower-mass systems need to come close to the host compared to the higher-mass satellites in order to trigger the SF burst. This is also evident in model \mymodels{48} where most of the pericentric distances are quite far from the centre of the host, and only the latest two passages~($19$ and $15$~kpc) correlate with the prominent short time-scale increase of the SF rate.

To summarize, in the case of pure tidal interaction between gasless satellites and massive MW-type disc galaxies, the SF increases in the hosts if the pericentric distance is smaller than $\approx 20$~kpc. At larger distances, we do not detect any significant effects, even in the case of massive~($\approx 6\times 10^{10}~\Msun$) perturber. 

\subsection{Models with gas-rich satellite galaxies}\label{sec::results_sfr_gas}
Next, we discuss the behaviour of the SFR in the simulations, which include a substantial amount of gas inside the satellite galaxies. In such models, we can follow the impact of the infall of gas stripped from the satellite galaxies and thus compare its contribution to the SFR of the host undergoing the purely kinetic perturbations in gasless simulations as discussed above. Remember that in this set of simulations, we use the same mass models of dwarf galaxies resulting in very similar orbits and pericentric distances over time~(see Fig.~\ref{fig::orbits}). Fig.~\ref{fig::sfr_gas} shows the orbital decay and SFH of the hosts compared to the isolated galaxy simulation. We notice that across all of the models with gas-rich satellites, the star formation rate is on average higher and the SFHs are significantly smoother when compared to the gasless satellite models and the isolated simulation. This is likely the result of the immediate interactions between the gas components of the host and satellite galaxy even at very early phases of evolution when the gas of the satellite is being stretched along the direction of the tidal forces~\citep[see also][]{2018MNRAS.478.5263T}. Furthermore, the satellite gas being stripped along the orbit is constantly infalling onto the host. This results in a negligible effect of close passages on the star formation rate, which is seen in Fig.~\ref{fig::sfr_gas}. In particular, although the star formation is much higher on average, we do not detect any prominent bursts in model \mymodels{41} which is the analogue of model \mymodels{47}, where we observed multiple SF bursts. In gas-rich satellite simulations, local peaks of the SF are being observed only in the case of massive satellites~(model \mymodels{42} and \mymodels{43}) at the very last stages of the merger when the satellite crosses the disc of the host. 

%%%%%%%%%%%%%%%%%%%%%%%%%%%%%%%%%%%%%%%%%%%%%%%%%%%%%%%%%%%%%%%
\begin{figure*}
\begin{center}
\includegraphics[width=0.5\hsize]{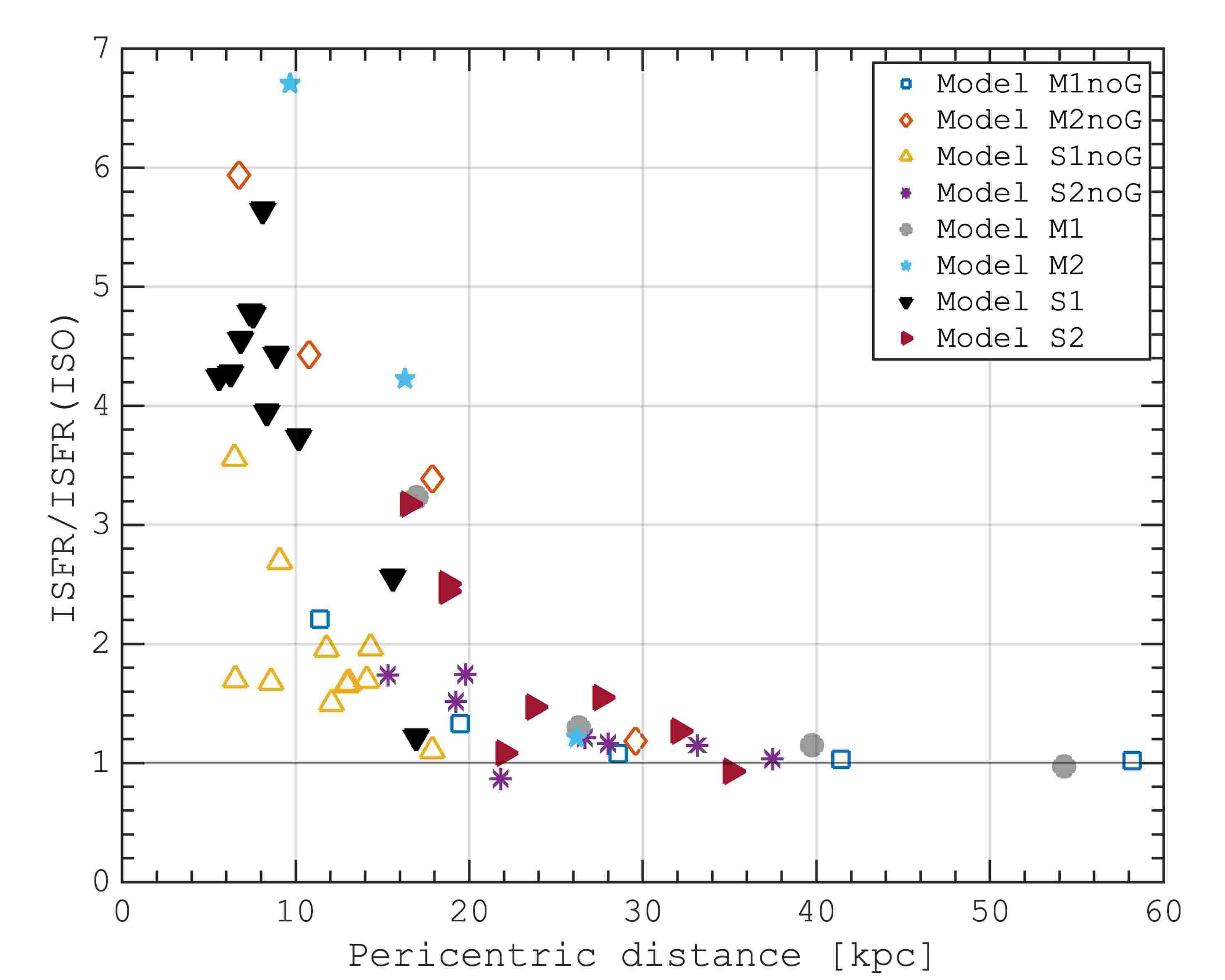}\includegraphics[width=0.5\hsize]{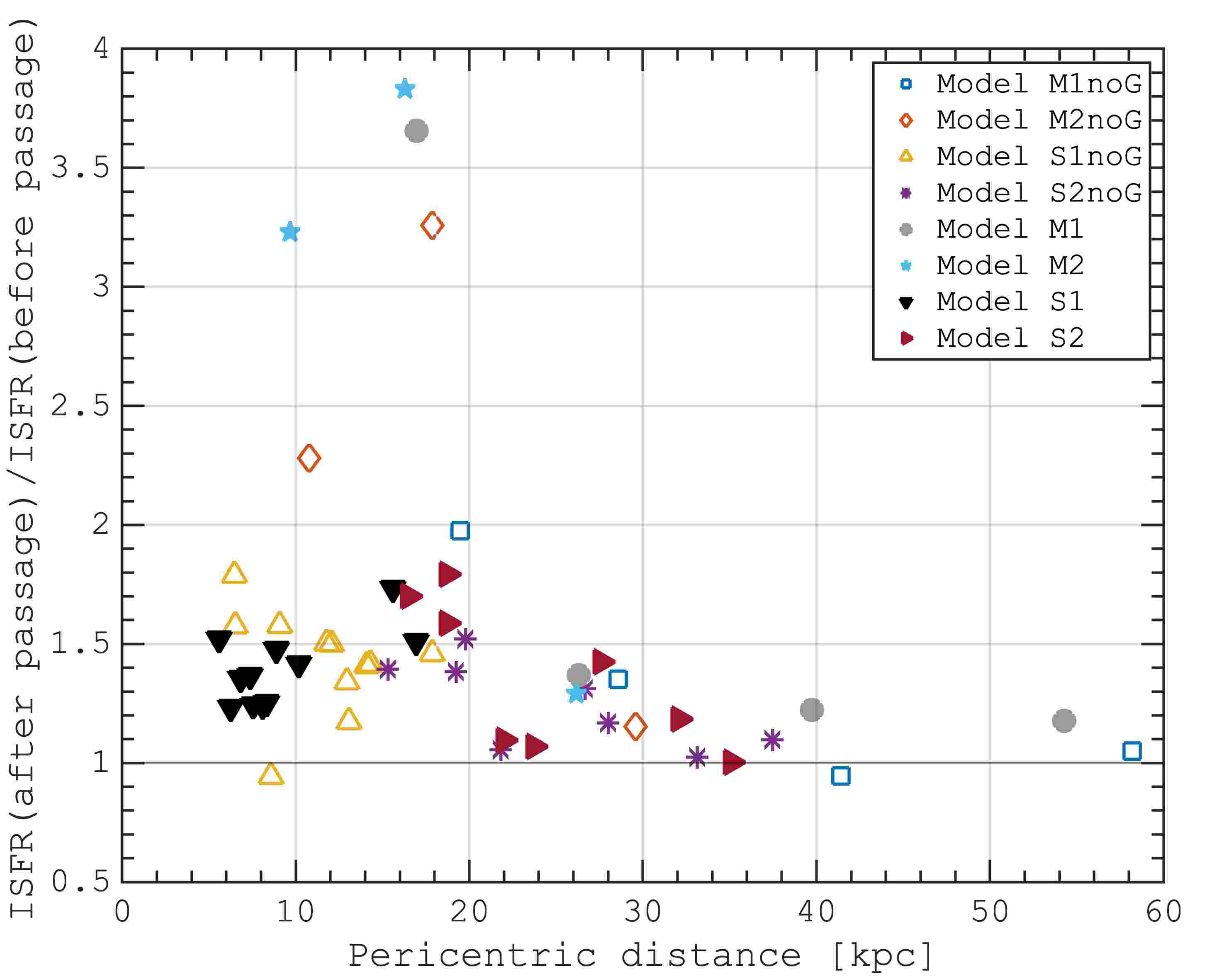}
\caption{Star formation rate enhancement as a function of pericentric distance. {\it In left:} the enhancement of the star formation is calculated as a ratio between the integrated star formation rate~(ISFR, see Eq.~\ref{eq::isfr}) in interacting and isolated simulations. In both isolated and interacting simulations, the integrated star formation rates are calculated over $0.5$~Gyr after the corresponding pericentric passage. Our simulations show a lack of enhancement of the SF unless the satellites pass closer than $\approx 20$~kpc in the case of $\leq6\times10^{10}\Msun$ satellite. {\it In right:} the enhancement of the star formation is calculated in interacting simulations only as a ratio between the integrated star formation rate before and after the pericentric passage of the satellite galaxy.}\label{fig::isfr}
\end{center}
\end{figure*}
%%%%%%%%%%%%%%%%%%%%%%%%%%%%%%%%%%%%%%%%%%%%%%%%%%%%%%%%%%%%%%%

To summarize, in the case of gas-rich satellite interactions, we observe a higher mean star formation rate compared to gasless satellite simulations. However, the pericentric passages do not cause any prominent bursts of star formation in the host galaxies. These results are in agreement with \cite{2021MNRAS.503.5846R} who found that the infall of gas onto the galaxy, at the early stages of the merger, is smooth, does not fuel the star-forming regions, and does not result in a rapid increase of the SF.

\subsection{SF enhancement due to interactions}\label{sec::results_sfr_where} % as a function of $\rm R_{gal}$}\label{sec::results_sfr_where}

We showed that the mean star formation rate is higher in models with orbiting satellites compared to the isolated galaxy and its higher even more if there is gas accretion from the satellites. However, which regions of the host galaxies are affected the most by the perturbations? In Fig.~\ref{fig::relative_sfr}, we demonstrate the degree to which star formation is enhanced in a given interacting model compared to the isolated galaxy simulation at different radii over time. The maps in the figure are colour-coded with the following parameters:
\begin{equation}
\rm \displaystyle \delta(R_{gal},t) = (\Sigma_{int}(R,t) - \Sigma_{iso}(R,t))/(\Sigma_{int}(R,t) + \Sigma_{iso}(R,t))\,,
\end{equation}
where $\rm \Sigma_{int}$ and $\rm \Sigma_{iso}$ are the star formation surface density in interacting and isolated simulations, respectively. The parameter $\delta(R_{gal},t)$ shows the degree to which the star formation rate in a pair of models is different. If it is negative, then the SFR is higher in the isolated galaxy; if it is close to zero, then the SFRs are comparable; and once it is close to $1$, then the SFR in the isolated galaxy is negligible compared to the isolated galaxy model. The blue vertical lines highlight the pericentric passages of satellites if they are within $19$~kpc. Otherwise, the arrows and numbers in blue boxes show the time and pericentric distance, respectively. 

%%%%%%%%%%%%%%%%%%%%%%%%%%%%%%%%%%%%%%%%%%%%%%%%%%%%%%%%%%%%%%%
\begin{figure*}
\begin{center}
\includegraphics[width=1\hsize]{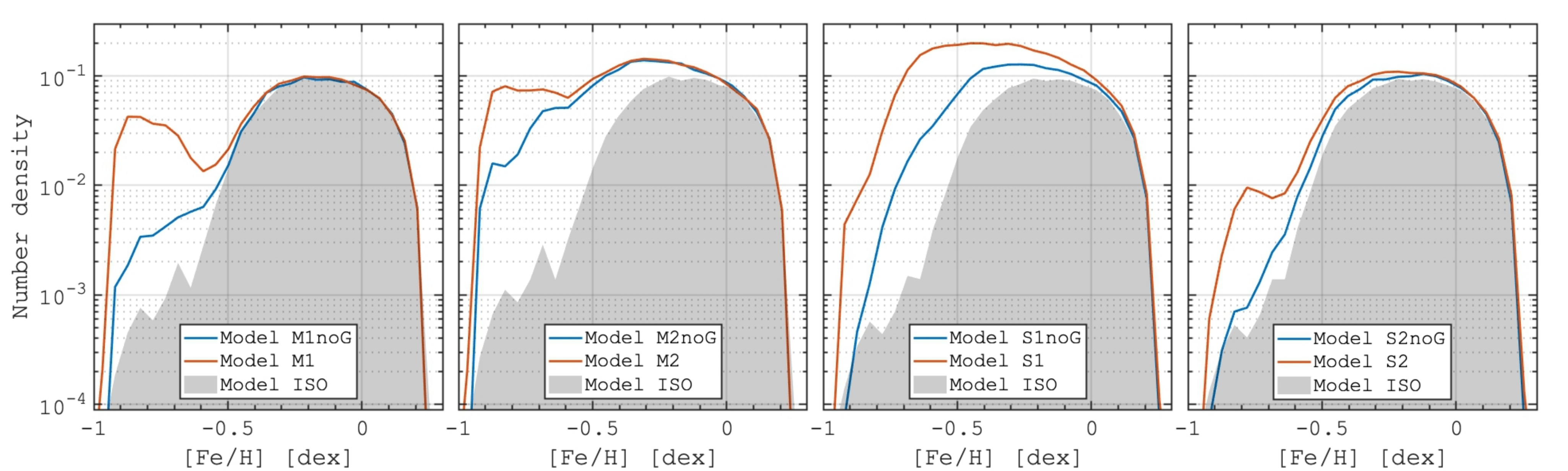}
\caption{Metallicity distribution functions~(MDFs) of stars formed inside the host galaxy. The grey area is the same across the panels and corresponds to the isolated galaxy simulation. Each panel shows a pair of simulations with the same orbit of the satellite galaxy where the MDFs are normalized by the total mass of stars formed in isolated simulation. The blue and red solid lines correspond to the model with a gasless satellite and a gas-rich satellite, respectively.}\label{fig::mdfs}
\end{center}
\end{figure*}
%%%%%%%%%%%%%%%%%%%%%%%%%%%%%%%%%%%%%%%%%%%%%%%%%%%%%%%%%%%%%%%

Fig.~\ref{fig::relative_sfr} clearly demonstrates that in our models, the SF is enhanced mostly in the outer regions of the host galaxy with a small variance in the innermost region compared to the isolated simulation. In addition, all our models suggest that the SF starts to increase at $\approx 10$~kpc once the satellites are sufficiently close to the host. Subsequently, the region of enhanced SF propagates slowly towards inner regions. However, even in models with strong impact~(models \mymodels{46}, \mymodels{47}, \mymodels{43} and \mymodels{41}), the inner regions show larger SF only several Gyr after the first close passage of the satellite. In these models, we also observe some SF out to $20$~kpc which is not the case in the isolated simulations with a weak perturbation from satellites.

The effect on the SF of the host we detected is strikingly similar to the kinematic response of the disc reported by \citep{2013MNRAS.429..159G}, where the MW disc hosts prominent vertical density features at $\rm R_{gal}>8$~kpc~(see their Fig. 6). \cite{2022MNRAS.516.5067C} also showed the impact of a massive satellite in the outer disc, where the radial migration strength increases with radius and the guiding radius increases by more than $1$~kpc explicitly where $\rm R_{gal}>8$~kpc~(see their Fig. 9). 

In Figure~\ref{fig::relative_sfr_maps}, we depict the proportional increase in gas surface density and star formation rate within the host galaxy plane, specifically focusing on simulation \mymodels{46}. This simulation is distinguished by the occurrence of two pericentric passages by a massive satellite before merging with the host. The upper panel illustrates the decay of the orbital trajectory, while the lower panels portray the instances immediately preceding, during, and immediately following each pericentric passage. The black crosses show the XY-position of the satellite galaxy at the corresponding times.

The enhancement, calculated as previously illustrated in Fig.~\ref{fig::relative_sfr}, is referenced relative to the isolated simulation by using snapshots taken at equivalent time points. The depicted figure highlights that during the initial passage, there is a commencement of perturbation in gas density, primarily observable in the outer regions. This results in a marginal increase of the SFR, evident at distances greater than $10$ kpc. A more recent and closer passage gives rise to prominent azimuthal variations in the parameters. Notably, the gas distribution exhibits pronounced tidal structures, concomitant with a corresponding increase in the SFR. However, caution is exercised in establishing a direct link between the disc region impacted by a satellite and the heightened SFR, owing to the fact that interactions induce tidal structures, which tend to wind up rapidly~\citep{2010MNRAS.403..625D,2016MNRAS.458.3990P}, thus averaging out their impact along the azimuth. Consequently, even if the interaction with the Sgr triggers localized enhancement of the SFR within the MW, particularly within a specific quadrant, the rapid evolution and subsequent dissipation of tidal structures render such effects undetectable. 

\subsection{Star formation enhancement as a function of pericentric distance}\label{sec::results_isfr}
In order to quantify the star formation enhancement for each pericentric passage, we calculate the integrated star formation rate, which is the total mass of stars formed over $0.5$~Gyr after a given pericentric passage~(see also \cite{2007A&A...468...61D}):  
\begin{equation}
\rm     ISFR = \int\displaylimits^{\tau+0.5 Gyr}_\tau SFR(t) dt\,\label{eq::isfr}
\end{equation}
where $\rm SFR(t)$ is the star formation rate, and $\tau$ is the time of pericentric passage of a satellite galaxy. We have chosen a time period of $0.5$~Gyr to capture a pure effect of the interaction and avoid a possible overlap between two subsequent pericentric passages. In Fig.~\ref{fig::isfr}, different models are shown by different markers and colours. On the left, for each pericentric passage, the ISFR is normalized by the ISFR in the isolated simulation calculated over the same period of time, and it is shown as a function of the pericentric distance. The figure illustrates the previously-described behaviour of the SF rate in all of our simulations. In particular, we do not detect any systematic increase of the SF until $30$~kpc; at closer passages~($20-30$~kpc), the SF intensity starts to increase, but it is still not larger by a factor of $2$, relative to the isolated model. Only once the satellite galaxies pass within $20$~kpc, the star formation is enhanced by a factor of $2-4$. For even closer pericenters, the SF rate enhancement is even higher, but the most massive mergers end up with the SF quenching; however, the core of the satellite moves inside the disc resembling the very latest stages of the merger. Since this has not yet occurred with the Sgr dwarf, this is beyond the interest of our work. In the right panel of Fig.~\ref{fig::isfr} we show the ratio between the ISFR after and before the pericentric passage in interacting simulations only. This allows us to quantify how much the SF changed due to interaction. It appears that only in a few models the SFR is increased by a factor of $> 3$, while in most of the cases, we detect an increase smaller than $2$.

%, which is not yet happened with the Sgr dwarf, and thus, beyond the interest of our work. 

\subsection{Metallicity variations caused by the interactions}\label{sec::results_metallicity}

A number of recent works focus on the effects of mergers in the metallicity distribution in simulated galaxies~\citep{2010A&A...518A..56M, 2010ApJ...710L.156R, 2012ApJ...746..108T, 2018MNRAS.479.3381B,2022MNRAS.509.2720S,2022arXiv220605491K}. These works clearly show the ISM metallicity dilution effect in the central regions of the galaxy due to metal-poor gas being tidally funneled into the centre during mergers. In agreement with the theoretical works, \cite{2020MNRAS.494.3469B} investigated a large sample of interacting galaxies from the SDSS DR7 and found an increasing gas-phase metallicity dilution and enhanced star formation activity with decreasing separation between galaxies~\citep[see also, e.g.][]{2015MNRAS.451.4005G}.

These works study the major mergers of massive galaxies, which were completed on a short time scale. However, the Sgr-MW interaction is less violent so far. In this section, we discuss the behaviour of the metallicity distribution in our simulations and some significant features caused by the perturbations from the satellites and corresponding gas infall onto the host galaxy. The particular details of the metallicity behaviour are dependent on the adopted initial metallicity; thus, we discuss only some general metallicity trends without giving any quantitative predictions regarding the metallicity distribution function~(MDF) features caused by the Sgr dwarf galaxy in the MW.

First, in Fig.~\ref{fig::mdfs} we show the metallicity distributions for the stars formed inside the host galaxy. Each panel shows the MDFs from a pair of simulations with the same satellite orbit with gasless~(blue) and gas-rich~(red) satellite which are compared to the MDF of the isolated host model~(gray area). A striking feature of all our interacting simulations is the presence of a certain fraction of relatively low-\FeH\ stars. In gasless satellite simulations~(blue lines), the excess of low-\FeH\ stars results from the enhanced stars formation in the outer disc~(see Fig.~\ref{fig::relative_sfr}) where the ISM metallicity is low because of the imposed negative gradient in the ISM of the host~(see Section~\ref{sec::modelICs} for details). We see that the only perturbation of the outer disc naturally leads to the formation of a non-negligible amount of stars with relatively low metallicity inherited from the ISM located in the outer disc. 

The impact of the gas infall from satellites is clearly seen once we compare a pair of models in each panel of Fig.~\ref{fig::mdfs}. First, the total mass of the low-\FeH\ stars is larger in gas-rich satellites simulations~(red lines) than in gasless satellite simulations~(blue lines), thus, suggesting that some of the stars were formed in-situ but in gas which either was directly accreted from satellites and/or mixed with the pre-existing gas of the host. Also, in most of the gas-rich satellite models, we see a second peak of the MDFs which is not observed in the gasless simulations. This further emphasizes the degree to which the gas from the satellite pollutes the host ISM and dilutes the metallicity of newly forming stars of the host, which is in agreement with~\cite{2016A&A...586A.112R}.

\begin{figure*}
\begin{center}
\includegraphics[width=1\hsize]{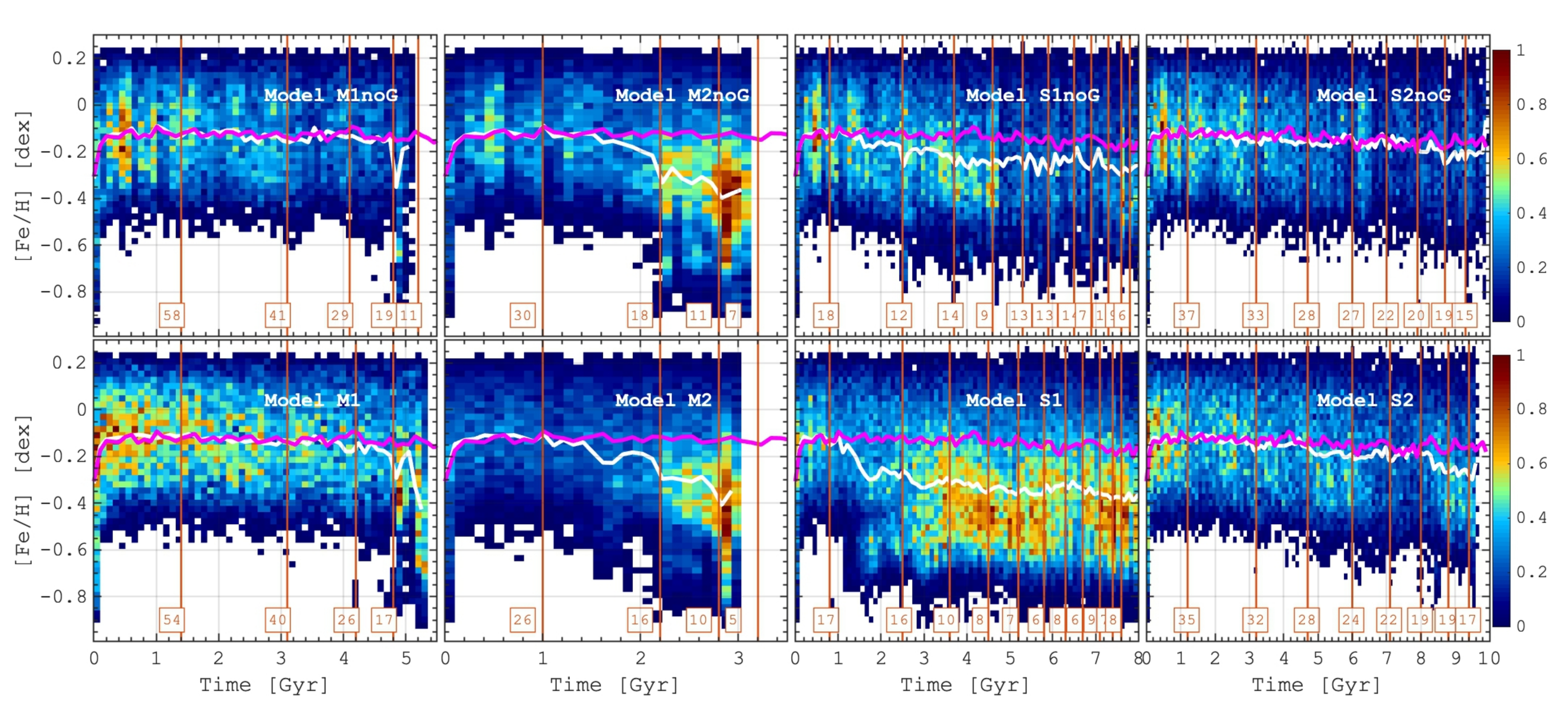}
\caption{Age - metallicity relation of stars formed inside the host galaxy. The density maps are normalized by the maximum value in each panel. The red vertical lines correspond to the moments of pericentric passages of satellites, where the distance is marked in corresponding red boxes. The solid white and magenta lines show the evolution of the mean metallicity in a given model and in the isolated simulation, respectively.}\label{fig::age_met}
\end{center}
\end{figure*}

To highlight the effect of the metallicity dilution, we show the age-metallicity relation for the stars formed inside the host galaxies in Fig.~\ref{fig::age_met}.
The ages of stars are indicative of their formation times in a manner where the youngest stars originate towards the end of the simulation, and the highest age corresponds to the simulation's duration. The density distributions depict the MDFs evolution as a function of time. The evolution of the mean metallicity in a particular model~(white lines) is compared to the one in the isolated simulation~(magenta lines). The effect of the metallicity dilution is inevitably linked to the close~($<20$~kpc) pericentric passages, where the mean metallicity decreases after the close approaches of the satellites. The dilution is more prominent in gas-rich satellite simulations because of the low-\FeH\ gas accretion onto the host galaxy. Note, however, that similar to the SFR behaviour~(see Figs.~\ref{fig::sfr_nogas},~\ref{fig::sfr_gas}), the mean metallicity steadily decreases after the first close passage and does not show a quick recovery to its previous level. Hence, our simulations are in agreement with some previous studies~(see, e.g. \cite{2010A&A...518A..56M} who showed that fly-bys of galaxies cause some dilution of the ISM metallicity). Moreover, some recent studies of the MW chemical compositions suggest a similar effect potentially associated to the Sgr and GSE accretion events~\citep{2023MNRAS.tmp.1561R,2023MNRAS.tmpL..32C,2023A&A...673A.155Q}.

%%%%%%%%%%%%%%%%%%%%%%%%%%%%%%%%%%%%%%%%%%%%%%%%%%%%%%%%%%%%%%%
%%%%%%%%%%%%%%%%%%%%%%%%%%%%%%%%%%%%%%%%%%%%%%%%%%%%%%%%%%%%%%%

\section{Discussion}\label{sec::discussion}
%Remind that in this paper, we do not aim to reproduce the exact impact of the Sgr dwarf galaxy on the MW evolution. However, using a set of models, which are qualitatively similar to the ones we expect for the Sgr-MW interactions,  we now discuss the implication of our models in the context of the MW past. 
Using a set of models that are qualitatively similar to the ones we expect for the Sgr-MW interactions,  we now discuss the implication of our models in the context of the MW evolution.

By using a CMD-fitting technique, \cite{2020NatAs...4..965R} found an excess of stellar populations in several age-bins which were interpreted as bursts of the star formation, at $\approx 1, 2$ and $6$~Gyr ago~(note that \cite{2006A&A...459..783C} and \cite{2021A&A...647A..39S} also found recent peaks of the SF in the SNd data). Although the age-distribution of stars in a local SNd-like region may not perfectly match the star formation history of the entire galaxy in part because of stellar radial migration~(Attard et al. in prep), it was suggested that the bursts of the SF are real and correlate with the pericentric passages of the Sgr dwarf galaxy. Most of the dynamical models of the Sgr orbit suggest a couple of recent passages~($\approx 1$ and $\approx 2$ Gyr ago) with pericentric distances of $\approx 10-20$~kpc~\citep[e.g.][]{2010ApJ...714..229L,2010ApJ...725.1516L,2020MNRAS.497.4162V,2022ApJ...927..131B}, and since the present-day core of the Sgr galaxy does not contain gas~\citep{2018MNRAS.478.5263T}, these two conditions met in our models~(see Fig.~\ref{fig::sfr_nogas}). Therefore, we confirm that the most recent passages of the Sgr dwarf galaxy indeed can have a significant impact on the SFH of the MW, similar to the one proposed by \cite{2020NatAs...4..965R}. Note, however, that our models suggest that, in this case, the SF should be enhanced more in the outer disc~($>10$~kpc, see Fig.~\ref{fig::relative_sfr}), whose magnitude, however, should depend on the previous passages of the Sgr.

The moments of previous passages~($>3$~Gyr ago) of the Sgr are less certain because of the initial mass and the mass-loss of the Sgr adopted in a given model. For instance, in the low-mass satellite simulations~\citep[$<2\times 10^{10}$~\Msun, e.g.][]{1996ApJ...465..278J,2010MNRAS.408L..26P,2021MNRAS.508.1459H}, we expect many pericentric passages~(nearly for each Gyr back in time) which apparently were not detected in the SFH~($2-5$~Gyr ago) of the MW by \cite{2020NatAs...4..965R}. 

However, a very large mass of the Sgr progenitor~($>4-6 \times 10^{10}$~\Msun) at its first infall has been proposed in a number of different studies~\citep[see, e.g.][]{2000MNRAS.314..468J,2017MNRAS.464..794G,2019MNRAS.487.5799R,2022MNRAS.516.1685D}. The models which reproduce the best the correlation between the SF burst and pericentric passages of the Sgr are presented in \cite{2018MNRAS.481..286L}, where we focus on the first encounter. According to their models~(L1, L2, H1 and H2) the Sgr's infall $\rm t_{infall}$ should have happened $4-6$~Gyr ago. Model L1 ($\rm t_{infall}\approx5.5$~Gyr ago) was already ruled out due to its inability to reproduce the outer structure of the MW~\citep{2018MNRAS.481..286L}. Models L2 and H2 show $\rm t_{infall}\approx4$~Gyr ago which is by $\rm \approx1.5$~Gyr off from the first and the most significant SF burst~($\approx 6$~Gyr) found by \cite{2020NatAs...4..965R}. At the same time, the remaining model H1~($\rm t_{infall}\approx5$~Gyr ago) marginally reproduces the Monoceros Ring~\citep{2018MNRAS.481..286L}. 

Nevertheless, one could imagine that some models with parameters similar to H1/H2 and L2 from \cite{2018MNRAS.481..286L} can simultaneously reproduce the outer MW disc features and predict the first pericentric passage exactly at the time expected from the MW SFH~($\approx 6$~Gyr ago). In this case, the first pericenter should be in the range of $30-40$~kpc from the MW centre~(see Fig. 2 in \cite{2018MNRAS.481..286L}), otherwise, the dynamical friction would have already dragged a still quite massive Sgr to the MW centre because of a rather moderate mass loss during pericentric passages~\citep[][]{2017MNRAS.471.4170T}. At the first infall, the Sgr should be still relatively gas-rich, which is supported by the star formation history of the galaxy~\citep{2007ApJ...667L..57S,2015MNRAS.451.3489D,2021ApJ...923..172H}. In this case, however, our models do not predict a coherent burst of star formation~(see Fig.~\ref{fig::isfr}). Therefore, in order to simultaneously account for a massive Sgr precursor at the first infall, its close first passage~($<20$~kpc), and to avoid its fast subsequent coalescence with the MW, we require a substantial mass loss during the first encounter~\citep[see, e.g.][]{1995ApJ...451..598J} which is possibly even more significant than the one suggested by \cite{2021MNRAS.504.3168B}. 

\section{Conclusions}\label{sec::summary}
In this work, we have conducted a series of hydrodynamical simulations of interactions of the MW-type galaxies with massive satellites on quasi-polar orbits. We studied the purely dynamical impact of gasless satellites and a more complex evolution in the case of gas-rich satellites. The aim of our models is to address whether pericentric passages of satellites cause short-time scale bursts of SF in the host galaxies. Although we found a substantial increase in the star formation rate in the host compared to isolated galaxy simulation, the details of the SFHs are highly dependent on the orbit and amount of gas in the satellite galaxies. Our main conclusions are summarized below.

\begin{itemize}
    \item Perturbations caused by the close passages of a massive satellite result in enhanced star formation inside the MW-type discs. Prominent and short-term bursts of star formation are the results of very close ($<20$~kpc) pericentric passages of massive ($(2-6)\times 10^{10}$~\Msun) satellites which do not contain gas~(see Figs.~\ref{fig::sfr_nogas} and~\ref{fig::isfr}). For a given pericentric distance, the effect on SF is stronger in the case of a more massive satellite. 
    \item Interactions with gas-rich satellites result in even higher SF rate enhancement, however, in this case, we do not detect prominent SF bursts unless the satellite passes through the disc of the host at the latest stages of coalescence. The reason for such behaviour is the steady accretion of gas from the satellite, which smoothens the overall SF rate evolution in the host~(see Fig.~\ref{fig::sfr_gas}).
    \item Independent of the gas content of the satellite, the SFR is mostly enhanced in the outer disc regions of the host after the first close~($<20$~kpc) pericentric passage. In particular, soon after the close pericentric passage, the SF is enhanced in the outer disc~($>10$~kpc), and then the SF propagates towards the centre of the host over the course of several Gyr. Therefore, we do not see an imminent impact on the inner SF of the host galaxy in case of close passages of gas-rich satellites~(see Fig.~\ref{fig::relative_sfr}).
    \item The increase of the SF in the outer disc of the host results in the excess of stars with lower-\FeH\, which is being inherited from the ISM at the outskirts~(see Figs.~\ref{fig::mdfs} and~\ref{fig::age_met}). The effect of stellar metallicity dilution is seen in both gasless and gas-rich satellite simulations, where it is more prominent in gas-rich satellite simulations due to low-\FeH\ gas accretion onto the host galaxy. 
    \item Our simulations are in favour of the causality between the latest passages of the Sgr galaxy and the bursts of the SF in the SNd at $\approx 1$ and $\approx 2$~Gyr~\citep{2020NatAs...4..965R,2021A&A...647A..39S}. However, according to our model, a potential burst at $\approx 6$~Gyr~\citep{2020NatAs...4..965R} requires a very close passage~($<20$~kpc from the MW centre) of a massive~($>2\times 10^{10}$~\Msun) Sgr precursor with a very substantial mass loss allowing to prevent a rapid coalescence with the MW until the present. 
\end{itemize}

\section*{Acknowledgements}
We sincerely thank the referee for a thoughtful review of our paper. BA thanks the AIP for the access to the institute's infrastructure during his remote internship. This work has made use of the computational resources available at the AIP, as well as those obtained through the DARI grant A1020410154 (PI: P. Di Matteo). SK thanks participants of weekly chemodynamics meetings at the AIP for their useful suggestions regarding the results of this work.

%%%%%%%%%%%%%%%%%%%%%%%%%%%%%%%%%%%%%%%%%%%%%%%%%%
\section*{Data Availability}
The data supporting this article will be shared upon reasonable request sent to the corresponding authors.



\bibliographystyle{mnras}
\bibliography{refs} % if your bibtex file is called example.bib

\begin{thebibliography}{}
\makeatletter
\relax
\def\mn@urlcharsother{\let\do\@makeother \do\$\do\&\do\#\do\^\do\_\do\%\do\~}
\def\mn@doi{\begingroup\mn@urlcharsother \@ifnextchar [ {\mn@doi@}
  {\mn@doi@[]}}
\def\mn@doi@[#1]#2{\def\@tempa{#1}\ifx\@tempa\@empty \href
  {http://dx.doi.org/#2} {doi:#2}\else \href {http://dx.doi.org/#2} {#1}\fi
  \endgroup}
\def\mn@eprint#1#2{\mn@eprint@#1:#2::\@nil}
\def\mn@eprint@arXiv#1{\href {http://arxiv.org/abs/#1} {{\tt arXiv:#1}}}
\def\mn@eprint@dblp#1{\href {http://dblp.uni-trier.de/rec/bibtex/#1.xml}
  {dblp:#1}}
\def\mn@eprint@#1:#2:#3:#4\@nil{\def\@tempa {#1}\def\@tempb {#2}\def\@tempc
  {#3}\ifx \@tempc \@empty \let \@tempc \@tempb \let \@tempb \@tempa \fi \ifx
  \@tempb \@empty \def\@tempb {arXiv}\fi \@ifundefined
  {mn@eprint@\@tempb}{\@tempb:\@tempc}{\expandafter \expandafter \csname
  mn@eprint@\@tempb\endcsname \expandafter{\@tempc}}}

\bibitem[\protect\citeauthoryear{{Antoja} et~al.,}{{Antoja}
  et~al.}{2018}]{2018Natur.561..360A}
{Antoja} T.,  et~al., 2018, \mn@doi [\nat] {10.1038/s41586-018-0510-7}, \href
  {https://ui.adsabs.harvard.edu/abs/2018Natur.561..360A} {561, 360}

\bibitem[\protect\citeauthoryear{{Bailin}}{{Bailin}}{2003}]{2003ApJ...583L..79B}
{Bailin} J.,  2003, \mn@doi [\apjl] {10.1086/368160}, \href
  {https://ui.adsabs.harvard.edu/abs/2003ApJ...583L..79B} {583, L79}

\bibitem[\protect\citeauthoryear{{Barnes} \& {Hernquist}}{{Barnes} \&
  {Hernquist}}{1991}]{1991ApJ...370L..65B}
{Barnes} J.~E.,  {Hernquist} L.~E.,  1991, \mn@doi [\apjl] {10.1086/185978},
  \href {https://ui.adsabs.harvard.edu/abs/1991ApJ...370L..65B} {370, L65}

\bibitem[\protect\citeauthoryear{{Bennett} \& {Bovy}}{{Bennett} \&
  {Bovy}}{2021}]{2021MNRAS.503..376B}
{Bennett} M.,  {Bovy} J.,  2021, \mn@doi [\mnras] {10.1093/mnras/stab524},
  \href {https://ui.adsabs.harvard.edu/abs/2021MNRAS.503..376B} {503, 376}

\bibitem[\protect\citeauthoryear{{Bennett}, {Bovy}  \& {Hunt}}{{Bennett}
  et~al.}{2022}]{2022ApJ...927..131B}
{Bennett} M.,  {Bovy} J.,   {Hunt} J. A.~S.,  2022, \mn@doi [\apj]
  {10.3847/1538-4357/ac5021}, \href
  {https://ui.adsabs.harvard.edu/abs/2022ApJ...927..131B} {927, 131}

\bibitem[\protect\citeauthoryear{{Bland-Hawthorn} \&
  {Gerhard}}{{Bland-Hawthorn} \& {Gerhard}}{2016}]{2016ARA&A..54..529B}
{Bland-Hawthorn} J.,  {Gerhard} O.,  2016, \mn@doi [\araa]
  {10.1146/annurev-astro-081915-023441}, \href
  {https://ui.adsabs.harvard.edu/abs/2016ARA&A..54..529B} {54, 529}

\bibitem[\protect\citeauthoryear{{Bland-Hawthorn} \&
  {Tepper-Garc{\'\i}a}}{{Bland-Hawthorn} \&
  {Tepper-Garc{\'\i}a}}{2021}]{2021MNRAS.504.3168B}
{Bland-Hawthorn} J.,  {Tepper-Garc{\'\i}a} T.,  2021, \mn@doi [\mnras]
  {10.1093/mnras/stab704}, \href
  {https://ui.adsabs.harvard.edu/abs/2021MNRAS.504.3168B} {504, 3168}

\bibitem[\protect\citeauthoryear{{Bland-Hawthorn} et~al.,}{{Bland-Hawthorn}
  et~al.}{2019}]{2019MNRAS.486.1167B}
{Bland-Hawthorn} J.,  et~al., 2019, \mn@doi [\mnras] {10.1093/mnras/stz217},
  \href {https://ui.adsabs.harvard.edu/abs/2019MNRAS.486.1167B} {486, 1167}

\bibitem[\protect\citeauthoryear{{Bustamante}, {Sparre}, {Springel}  \&
  {Grand}}{{Bustamante} et~al.}{2018}]{2018MNRAS.479.3381B}
{Bustamante} S.,  {Sparre} M.,  {Springel} V.,   {Grand} R. J.~J.,  2018,
  \mn@doi [\mnras] {10.1093/mnras/sty1692}, \href
  {https://ui.adsabs.harvard.edu/abs/2018MNRAS.479.3381B} {479, 3381}

\bibitem[\protect\citeauthoryear{{Bustamante}, {Ellison}, {Patton}  \&
  {Sparre}}{{Bustamante} et~al.}{2020}]{2020MNRAS.494.3469B}
{Bustamante} S.,  {Ellison} S.~L.,  {Patton} D.~R.,   {Sparre} M.,  2020,
  \mn@doi [\mnras] {10.1093/mnras/staa1025}, \href
  {https://ui.adsabs.harvard.edu/abs/2020MNRAS.494.3469B} {494, 3469}

\bibitem[\protect\citeauthoryear{{Carr}, {Johnston}, {Laporte}  \&
  {Ness}}{{Carr} et~al.}{2022}]{2022MNRAS.516.5067C}
{Carr} C.,  {Johnston} K.~V.,  {Laporte} C. F.~P.,   {Ness} M.~K.,  2022,
  \mn@doi [\mnras] {10.1093/mnras/stac2403}, \href
  {https://ui.adsabs.harvard.edu/abs/2022MNRAS.516.5067C} {516, 5067}

\bibitem[\protect\citeauthoryear{{Cignoni}, {Degl'Innocenti}, {Prada Moroni}
  \& {Shore}}{{Cignoni} et~al.}{2006}]{2006A&A...459..783C}
{Cignoni} M.,  {Degl'Innocenti} S.,  {Prada Moroni} P.~G.,   {Shore} S.~N.,
  2006, \mn@doi [\aap] {10.1051/0004-6361:20065645}, \href
  {https://ui.adsabs.harvard.edu/abs/2006A&A...459..783C} {459, 783}

\bibitem[\protect\citeauthoryear{{Ciuc{\u{a}}} et~al.,}{{Ciuc{\u{a}}}
  et~al.}{2023}]{2023MNRAS.tmpL..32C}
{Ciuc{\u{a}}} I.,  et~al., 2023, \mn@doi [\mnras] {10.1093/mnrasl/slad033},
  \href {https://ui.adsabs.harvard.edu/abs/2023MNRAS.tmpL..32C} {}

\bibitem[\protect\citeauthoryear{{Cox}, {Jonsson}, {Somerville}, {Primack}  \&
  {Dekel}}{{Cox} et~al.}{2008}]{2008MNRAS.384..386C}
{Cox} T.~J.,  {Jonsson} P.,  {Somerville} R.~S.,  {Primack} J.~R.,   {Dekel}
  A.,  2008, \mn@doi [\mnras] {10.1111/j.1365-2966.2007.12730.x}, \href
  {https://ui.adsabs.harvard.edu/abs/2008MNRAS.384..386C} {384, 386}

\bibitem[\protect\citeauthoryear{{D'Onghia}, {Madau}, {Vera-Ciro}, {Quillen}
  \& {Hernquist}}{{D'Onghia} et~al.}{2016}]{2016ApJ...823....4D}
{D'Onghia} E.,  {Madau} P.,  {Vera-Ciro} C.,  {Quillen} A.,   {Hernquist} L.,
  2016, \mn@doi [\apj] {10.3847/0004-637X/823/1/4}, \href
  {https://ui.adsabs.harvard.edu/abs/2016ApJ...823....4D} {823, 4}

\bibitem[\protect\citeauthoryear{{D'Souza} \& {Bell}}{{D'Souza} \&
  {Bell}}{2022}]{2022MNRAS.512..739D}
{D'Souza} R.,  {Bell} E.~F.,  2022, \mn@doi [\mnras] {10.1093/mnras/stac404},
  \href {https://ui.adsabs.harvard.edu/abs/2022MNRAS.512..739D} {512, 739}

\bibitem[\protect\citeauthoryear{{Darling} \& {Widrow}}{{Darling} \&
  {Widrow}}{2019}]{2019MNRAS.484.1050D}
{Darling} K.,  {Widrow} L.~M.,  2019, \mn@doi [\mnras] {10.1093/mnras/sty3508},
  \href {https://ui.adsabs.harvard.edu/abs/2019MNRAS.484.1050D} {484, 1050}

\bibitem[\protect\citeauthoryear{{Di Cintio}, {Mostoghiu}, {Knebe}  \&
  {Navarro}}{{Di Cintio} et~al.}{2021}]{2021MNRAS.506..531D}
{Di Cintio} A.,  {Mostoghiu} R.,  {Knebe} A.,   {Navarro} J.~F.,  2021, \mn@doi
  [\mnras] {10.1093/mnras/stab1682}, \href
  {https://ui.adsabs.harvard.edu/abs/2021MNRAS.506..531D} {506, 531}

\bibitem[\protect\citeauthoryear{{Di Matteo}, {Combes}, {Melchior}  \&
  {Semelin}}{{Di Matteo} et~al.}{2007}]{2007A&A...468...61D}
{Di Matteo} P.,  {Combes} F.,  {Melchior} A.~L.,   {Semelin} B.,  2007, \mn@doi
  [\aap] {10.1051/0004-6361:20066959}, \href
  {https://ui.adsabs.harvard.edu/abs/2007A&A...468...61D} {468, 61}

\bibitem[\protect\citeauthoryear{{Di Matteo}, {Bournaud}, {Martig}, {Combes},
  {Melchior}  \& {Semelin}}{{Di Matteo} et~al.}{2008}]{2008A&A...492...31D}
{Di Matteo} P.,  {Bournaud} F.,  {Martig} M.,  {Combes} F.,  {Melchior} A.~L.,
   {Semelin} B.,  2008, \mn@doi [\aap] {10.1051/0004-6361:200809480}, \href
  {https://ui.adsabs.harvard.edu/abs/2008A&A...492...31D} {492, 31}

\bibitem[\protect\citeauthoryear{{Dierickx} \& {Loeb}}{{Dierickx} \&
  {Loeb}}{2017}]{2017ApJ...836...92D}
{Dierickx} M. I.~P.,  {Loeb} A.,  2017, \mn@doi [\apj]
  {10.3847/1538-4357/836/1/92}, \href
  {https://ui.adsabs.harvard.edu/abs/2017ApJ...836...92D} {836, 92}

\bibitem[\protect\citeauthoryear{{Dillamore}, {Belokurov}, {Evans}  \&
  {Price-Whelan}}{{Dillamore} et~al.}{2022}]{2022MNRAS.516.1685D}
{Dillamore} A.~M.,  {Belokurov} V.,  {Evans} N.~W.,   {Price-Whelan} A.~M.,
  2022, \mn@doi [\mnras] {10.1093/mnras/stac2311}, \href
  {https://ui.adsabs.harvard.edu/abs/2022MNRAS.516.1685D} {516, 1685}

\bibitem[\protect\citeauthoryear{{Dobbs}, {Theis}, {Pringle}  \&
  {Bate}}{{Dobbs} et~al.}{2010}]{2010MNRAS.403..625D}
{Dobbs} C.~L.,  {Theis} C.,  {Pringle} J.~E.,   {Bate} M.~R.,  2010, \mn@doi
  [\mnras] {10.1111/j.1365-2966.2009.16161.x}, \href
  {https://ui.adsabs.harvard.edu/abs/2010MNRAS.403..625D} {403, 625}

\bibitem[\protect\citeauthoryear{{Drimmel} \& {Spergel}}{{Drimmel} \&
  {Spergel}}{2001}]{2001ApJ...556..181D}
{Drimmel} R.,  {Spergel} D.~N.,  2001, \mn@doi [\apj] {10.1086/321556}, \href
  {https://ui.adsabs.harvard.edu/abs/2001ApJ...556..181D} {556, 181}

\bibitem[\protect\citeauthoryear{{Dwek} et~al.,}{{Dwek}
  et~al.}{1995}]{1995ApJ...445..716D}
{Dwek} E.,  et~al., 1995, \mn@doi [\apj] {10.1086/175734}, \href
  {https://ui.adsabs.harvard.edu/abs/1995ApJ...445..716D} {445, 716}

\bibitem[\protect\citeauthoryear{{Gaia Collaboration} et~al.,}{{Gaia
  Collaboration} et~al.}{2018}]{2018A&A...616A..10G}
{Gaia Collaboration} et~al., 2018, \mn@doi [\aap]
  {10.1051/0004-6361/201832843}, \href
  {https://ui.adsabs.harvard.edu/abs/2018A&A...616A..10G} {616, A10}

\bibitem[\protect\citeauthoryear{{Gibbons}, {Belokurov}  \& {Evans}}{{Gibbons}
  et~al.}{2017}]{2017MNRAS.464..794G}
{Gibbons} S.~L.~J.,  {Belokurov} V.,   {Evans} N.~W.,  2017, \mn@doi [\mnras]
  {10.1093/mnras/stw2328}, \href
  {https://ui.adsabs.harvard.edu/abs/2017MNRAS.464..794G} {464, 794}

\bibitem[\protect\citeauthoryear{{G{\'o}mez}, {Minchev}, {O'Shea}, {Beers},
  {Bullock}  \& {Purcell}}{{G{\'o}mez} et~al.}{2013}]{2013MNRAS.429..159G}
{G{\'o}mez} F.~A.,  {Minchev} I.,  {O'Shea} B.~W.,  {Beers} T.~C.,  {Bullock}
  J.~S.,   {Purcell} C.~W.,  2013, \mn@doi [\mnras] {10.1093/mnras/sts327},
  \href {https://ui.adsabs.harvard.edu/abs/2013MNRAS.429..159G} {429, 159}

\bibitem[\protect\citeauthoryear{{G{\'o}mez}, {Besla}, {Carpintero},
  {Villalobos}, {O'Shea}  \& {Bell}}{{G{\'o}mez}
  et~al.}{2015}]{2015ApJ...802..128G}
{G{\'o}mez} F.~A.,  {Besla} G.,  {Carpintero} D.~D.,  {Villalobos} {\'A}.,
  {O'Shea} B.~W.,   {Bell} E.~F.,  2015, \mn@doi [\apj]
  {10.1088/0004-637X/802/2/128}, \href
  {https://ui.adsabs.harvard.edu/abs/2015ApJ...802..128G} {802, 128}

\bibitem[\protect\citeauthoryear{{Gr{\o}nnow}, {Finlator}  \&
  {Christensen}}{{Gr{\o}nnow} et~al.}{2015}]{2015MNRAS.451.4005G}
{Gr{\o}nnow} A.~E.,  {Finlator} K.,   {Christensen} L.,  2015, \mn@doi [\mnras]
  {10.1093/mnras/stv1232}, \href
  {https://ui.adsabs.harvard.edu/abs/2015MNRAS.451.4005G} {451, 4005}

\bibitem[\protect\citeauthoryear{{Hasselquist} et~al.,}{{Hasselquist}
  et~al.}{2021}]{2021ApJ...923..172H}
{Hasselquist} S.,  et~al., 2021, \mn@doi [\apj] {10.3847/1538-4357/ac25f9},
  \href {https://ui.adsabs.harvard.edu/abs/2021ApJ...923..172H} {923, 172}

\bibitem[\protect\citeauthoryear{{Hopkins}}{{Hopkins}}{2015}]{2015MNRAS.450...53H}
{Hopkins} P.~F.,  2015, \mn@doi [\mnras] {10.1093/mnras/stv195}, \href
  {https://ui.adsabs.harvard.edu/abs/2015MNRAS.450...53H} {450, 53}

\bibitem[\protect\citeauthoryear{{Hopkins}}{{Hopkins}}{2017}]{2017arXiv171201294H}
{Hopkins} P.~F.,  2017, arXiv e-prints, \href
  {https://ui.adsabs.harvard.edu/abs/2017arXiv171201294H} {p. arXiv:1712.01294}

\bibitem[\protect\citeauthoryear{{Hopkins}, {Cox}, {Hernquist}, {Narayanan},
  {Hayward}  \& {Murray}}{{Hopkins} et~al.}{2013a}]{2013MNRAS.430.1901H}
{Hopkins} P.~F.,  {Cox} T.~J.,  {Hernquist} L.,  {Narayanan} D.,  {Hayward}
  C.~C.,   {Murray} N.,  2013a, \mn@doi [\mnras] {10.1093/mnras/stt017}, \href
  {https://ui.adsabs.harvard.edu/abs/2013MNRAS.430.1901H} {430, 1901}

\bibitem[\protect\citeauthoryear{{Hopkins}, {Narayanan}  \& {Murray}}{{Hopkins}
  et~al.}{2013b}]{2013MNRAS.432.2647H}
{Hopkins} P.~F.,  {Narayanan} D.,   {Murray} N.,  2013b, \mn@doi [\mnras]
  {10.1093/mnras/stt723}, \href
  {https://ui.adsabs.harvard.edu/abs/2013MNRAS.432.2647H} {432, 2647}

\bibitem[\protect\citeauthoryear{{Hunt}, {Stelea}, {Johnston}, {Gandhi},
  {Laporte}  \& {B{\'e}dorf}}{{Hunt} et~al.}{2021}]{2021MNRAS.508.1459H}
{Hunt} J. A.~S.,  {Stelea} I.~A.,  {Johnston} K.~V.,  {Gandhi} S.~S.,
  {Laporte} C. F.~P.,   {B{\'e}dorf} J.,  2021, \mn@doi [\mnras]
  {10.1093/mnras/stab2580}, \href
  {https://ui.adsabs.harvard.edu/abs/2021MNRAS.508.1459H} {508, 1459}

\bibitem[\protect\citeauthoryear{{Ibata} \& {Razoumov}}{{Ibata} \&
  {Razoumov}}{1998}]{1998A&A...336..130I}
{Ibata} R.~A.,  {Razoumov} A.~O.,  1998, \aap, \href
  {https://ui.adsabs.harvard.edu/abs/1998A&A...336..130I} {336, 130}

\bibitem[\protect\citeauthoryear{{Ibata}, {Gilmore}  \& {Irwin}}{{Ibata}
  et~al.}{1994}]{1994Natur.370..194I}
{Ibata} R.~A.,  {Gilmore} G.,   {Irwin} M.~J.,  1994, \mn@doi [\nat]
  {10.1038/370194a0}, \href
  {https://ui.adsabs.harvard.edu/abs/1994Natur.370..194I} {370, 194}

\bibitem[\protect\citeauthoryear{{Ibata}, {Gilmore}  \& {Irwin}}{{Ibata}
  et~al.}{1995}]{1995MNRAS.277..781I}
{Ibata} R.~A.,  {Gilmore} G.,   {Irwin} M.~J.,  1995, \mn@doi [\mnras]
  {10.1093/mnras/277.3.781}, \href
  {https://ui.adsabs.harvard.edu/abs/1995MNRAS.277..781I} {277, 781}

\bibitem[\protect\citeauthoryear{{Jiang} \& {Binney}}{{Jiang} \&
  {Binney}}{2000}]{2000MNRAS.314..468J}
{Jiang} I.-G.,  {Binney} J.,  2000, \mn@doi [\mnras]
  {10.1046/j.1365-8711.2000.03311.x}, \href
  {https://ui.adsabs.harvard.edu/abs/2000MNRAS.314..468J} {314, 468}

\bibitem[\protect\citeauthoryear{{Jog} \& {Solomon}}{{Jog} \&
  {Solomon}}{1992}]{1992ApJ...387..152J}
{Jog} C.~J.,  {Solomon} P.~M.,  1992, \mn@doi [\apj] {10.1086/171067}, \href
  {https://ui.adsabs.harvard.edu/abs/1992ApJ...387..152J} {387, 152}

\bibitem[\protect\citeauthoryear{{Johnston}, {Spergel}  \&
  {Hernquist}}{{Johnston} et~al.}{1995}]{1995ApJ...451..598J}
{Johnston} K.~V.,  {Spergel} D.~N.,   {Hernquist} L.,  1995, \mn@doi [\apj]
  {10.1086/176247}, \href
  {https://ui.adsabs.harvard.edu/abs/1995ApJ...451..598J} {451, 598}

\bibitem[\protect\citeauthoryear{{Johnston}, {Hernquist}  \&
  {Bolte}}{{Johnston} et~al.}{1996}]{1996ApJ...465..278J}
{Johnston} K.~V.,  {Hernquist} L.,   {Bolte} M.,  1996, \mn@doi [\apj]
  {10.1086/177418}, \href
  {https://ui.adsabs.harvard.edu/abs/1996ApJ...465..278J} {465, 278}

\bibitem[\protect\citeauthoryear{{Kerr}}{{Kerr}}{1957}]{1957AJ.....62...93K}
{Kerr} F.~J.,  1957, \mn@doi [\aj] {10.1086/107466}, \href
  {https://ui.adsabs.harvard.edu/abs/1957AJ.....62...93K} {62, 93}

\bibitem[\protect\citeauthoryear{{Khoperskov}, {Di Matteo}, {Gerhard}, {Katz},
  {Haywood}, {Combes}, {Berczik}  \& {Gomez}}{{Khoperskov}
  et~al.}{2019}]{2019A&A...622L...6K}
{Khoperskov} S.,  {Di Matteo} P.,  {Gerhard} O.,  {Katz} D.,  {Haywood} M.,
  {Combes} F.,  {Berczik} P.,   {Gomez} A.,  2019, \mn@doi [\aap]
  {10.1051/0004-6361/201834707}, \href
  {https://ui.adsabs.harvard.edu/abs/2019A&A...622L...6K} {622, L6}

\bibitem[\protect\citeauthoryear{{Khoperskov} et~al.,}{{Khoperskov}
  et~al.}{2023a}]{2022arXiv220604521K}
{Khoperskov} S.,  et~al., 2023a, \mn@doi [\aap] {10.1051/0004-6361/202244232},
  \href {https://ui.adsabs.harvard.edu/abs/2023A&A...677A..89K} {677, A89}

\bibitem[\protect\citeauthoryear{{Khoperskov} et~al.,}{{Khoperskov}
  et~al.}{2023b}]{2022arXiv220605491K}
{Khoperskov} S.,  et~al., 2023b, \mn@doi [\aap] {10.1051/0004-6361/202244234},
  \href {https://ui.adsabs.harvard.edu/abs/2023A&A...677A..91K} {677, A91}

\bibitem[\protect\citeauthoryear{{Laporte}, {G{\'o}mez}, {Besla}, {Johnston}
  \& {Garavito-Camargo}}{{Laporte} et~al.}{2018a}]{2018MNRAS.473.1218L}
{Laporte} C. F.~P.,  {G{\'o}mez} F.~A.,  {Besla} G.,  {Johnston} K.~V.,
  {Garavito-Camargo} N.,  2018a, \mn@doi [\mnras] {10.1093/mnras/stx2146},
  \href {https://ui.adsabs.harvard.edu/abs/2018MNRAS.473.1218L} {473, 1218}

\bibitem[\protect\citeauthoryear{{Laporte}, {Johnston}, {G{\'o}mez},
  {Garavito-Camargo}  \& {Besla}}{{Laporte}
  et~al.}{2018b}]{2018MNRAS.481..286L}
{Laporte} C. F.~P.,  {Johnston} K.~V.,  {G{\'o}mez} F.~A.,  {Garavito-Camargo}
  N.,   {Besla} G.,  2018b, \mn@doi [\mnras] {10.1093/mnras/sty1574}, \href
  {https://ui.adsabs.harvard.edu/abs/2018MNRAS.481..286L} {481, 286}

\bibitem[\protect\citeauthoryear{{Laporte}, {Minchev}, {Johnston}  \&
  {G{\'o}mez}}{{Laporte} et~al.}{2019}]{2019MNRAS.485.3134L}
{Laporte} C. F.~P.,  {Minchev} I.,  {Johnston} K.~V.,   {G{\'o}mez} F.~A.,
  2019, \mn@doi [\mnras] {10.1093/mnras/stz583}, \href
  {https://ui.adsabs.harvard.edu/abs/2019MNRAS.485.3134L} {485, 3134}

\bibitem[\protect\citeauthoryear{{Law} \& {Majewski}}{{Law} \&
  {Majewski}}{2010}]{2010ApJ...714..229L}
{Law} D.~R.,  {Majewski} S.~R.,  2010, \mn@doi [\apj]
  {10.1088/0004-637X/714/1/229}, \href
  {https://ui.adsabs.harvard.edu/abs/2010ApJ...714..229L} {714, 229}

\bibitem[\protect\citeauthoryear{{Layden} \& {Sarajedini}}{{Layden} \&
  {Sarajedini}}{2000}]{2000AJ....119.1760L}
{Layden} A.~C.,  {Sarajedini} A.,  2000, \mn@doi [\aj] {10.1086/301293}, \href
  {https://ui.adsabs.harvard.edu/abs/2000AJ....119.1760L} {119, 1760}

\bibitem[\protect\citeauthoryear{{{\L}okas}, {Kazantzidis}, {Majewski}, {Law},
  {Mayer}  \& {Frinchaboy}}{{{\L}okas} et~al.}{2010}]{2010ApJ...725.1516L}
{{\L}okas} E.~L.,  {Kazantzidis} S.,  {Majewski} S.~R.,  {Law} D.~R.,  {Mayer}
  L.,   {Frinchaboy} P.~M.,  2010, \mn@doi [\apj]
  {10.1088/0004-637X/725/2/1516}, \href
  {https://ui.adsabs.harvard.edu/abs/2010ApJ...725.1516L} {725, 1516}

\bibitem[\protect\citeauthoryear{{Lu}, {Minchev}, {Buck}, {Khoperskov},
  {Steinmetz}, {Libeskind}, {Cescutti}  \& {Freeman}}{{Lu}
  et~al.}{2022}]{2022arXiv221204515L}
{Lu} Y.,  {Minchev} I.,  {Buck} T.,  {Khoperskov} S.,  {Steinmetz} M.,
  {Libeskind} N.,  {Cescutti} G.,   {Freeman} K.~C.,  2022, \mn@doi [arXiv
  e-prints] {10.48550/arXiv.2212.04515}, \href
  {https://ui.adsabs.harvard.edu/abs/2022arXiv221204515L} {p. arXiv:2212.04515}

\bibitem[\protect\citeauthoryear{{Massana} et~al.,}{{Massana}
  et~al.}{2022}]{2022MNRAS.513L..40M}
{Massana} P.,  et~al., 2022, \mn@doi [\mnras] {10.1093/mnrasl/slac030}, \href
  {https://ui.adsabs.harvard.edu/abs/2022MNRAS.513L..40M} {513, L40}

\bibitem[\protect\citeauthoryear{{Michel-Dansac}, {Abadi}, {Navarro}  \&
  {Steinmetz}}{{Michel-Dansac} et~al.}{2011}]{2011MNRAS.414L...1M}
{Michel-Dansac} L.,  {Abadi} M.~G.,  {Navarro} J.~F.,   {Steinmetz} M.,  2011,
  \mn@doi [\mnras] {10.1111/j.1745-3933.2011.01035.x}, \href
  {https://ui.adsabs.harvard.edu/abs/2011MNRAS.414L...1M} {414, L1}

\bibitem[\protect\citeauthoryear{{Mihos} \& {Hernquist}}{{Mihos} \&
  {Hernquist}}{1994}]{1994ApJ...425L..13M}
{Mihos} J.~C.,  {Hernquist} L.,  1994, \mn@doi [\apjl] {10.1086/187299}, \href
  {https://ui.adsabs.harvard.edu/abs/1994ApJ...425L..13M} {425, L13}

\bibitem[\protect\citeauthoryear{{Miyamoto} \& {Nagai}}{{Miyamoto} \&
  {Nagai}}{1975}]{1975PASJ...27..533M}
{Miyamoto} M.,  {Nagai} R.,  1975, \pasj, \href
  {https://ui.adsabs.harvard.edu/abs/1975PASJ...27..533M} {27, 533}

\bibitem[\protect\citeauthoryear{{Montuori}, {Di Matteo}, {Lehnert}, {Combes}
  \& {Semelin}}{{Montuori} et~al.}{2010}]{2010A&A...518A..56M}
{Montuori} M.,  {Di Matteo} P.,  {Lehnert} M.~D.,  {Combes} F.,   {Semelin} B.,
   2010, \mn@doi [\aap] {10.1051/0004-6361/201014304}, \href
  {https://ui.adsabs.harvard.edu/abs/2010A&A...518A..56M} {518, A56}

\bibitem[\protect\citeauthoryear{{Moreno}, {Torrey}, {Ellison}, {Patton},
  {Bluck}, {Bansal}  \& {Hernquist}}{{Moreno}
  et~al.}{2015}]{2015MNRAS.448.1107M}
{Moreno} J.,  {Torrey} P.,  {Ellison} S.~L.,  {Patton} D.~R.,  {Bluck} A.
  F.~L.,  {Bansal} G.,   {Hernquist} L.,  2015, \mn@doi [\mnras]
  {10.1093/mnras/stv094}, \href
  {https://ui.adsabs.harvard.edu/abs/2015MNRAS.448.1107M} {448, 1107}

\bibitem[\protect\citeauthoryear{{Ness} \& {Lang}}{{Ness} \&
  {Lang}}{2016}]{2016AJ....152...14N}
{Ness} M.,  {Lang} D.,  2016, \mn@doi [\aj] {10.3847/0004-6256/152/1/14}, \href
  {https://ui.adsabs.harvard.edu/abs/2016AJ....152...14N} {152, 14}

\bibitem[\protect\citeauthoryear{{Pasetto}, {Grebel}, {Berczik}, {Chiosi}  \&
  {Spurzem}}{{Pasetto} et~al.}{2011}]{2011A&A...525A..99P}
{Pasetto} S.,  {Grebel} E.~K.,  {Berczik} P.,  {Chiosi} C.,   {Spurzem} R.,
  2011, \mn@doi [\aap] {10.1051/0004-6361/200913415}, \href
  {https://ui.adsabs.harvard.edu/abs/2011A&A...525A..99P} {525, A99}

\bibitem[\protect\citeauthoryear{{Pe{\~n}arrubia}, {Belokurov}, {Evans},
  {Mart{\'\i}nez-Delgado}, {Gilmore}, {Irwin}, {Niederste-Ostholt}  \&
  {Zucker}}{{Pe{\~n}arrubia} et~al.}{2010}]{2010MNRAS.408L..26P}
{Pe{\~n}arrubia} J.,  {Belokurov} V.,  {Evans} N.~W.,  {Mart{\'\i}nez-Delgado}
  D.,  {Gilmore} G.,  {Irwin} M.,  {Niederste-Ostholt} M.,   {Zucker} D.~B.,
  2010, \mn@doi [\mnras] {10.1111/j.1745-3933.2010.00921.x}, \href
  {https://ui.adsabs.harvard.edu/abs/2010MNRAS.408L..26P} {408, L26}

\bibitem[\protect\citeauthoryear{{Pettitt}, {Tasker}  \& {Wadsley}}{{Pettitt}
  et~al.}{2016}]{2016MNRAS.458.3990P}
{Pettitt} A.~R.,  {Tasker} E.~J.,   {Wadsley} J.~W.,  2016, \mn@doi [\mnras]
  {10.1093/mnras/stw588}, \href
  {https://ui.adsabs.harvard.edu/abs/2016MNRAS.458.3990P} {458, 3990}

\bibitem[\protect\citeauthoryear{{Pettitt}, {Tasker}, {Wadsley}, {Keller}  \&
  {Benincasa}}{{Pettitt} et~al.}{2017}]{2017MNRAS.468.4189P}
{Pettitt} A.~R.,  {Tasker} E.~J.,  {Wadsley} J.~W.,  {Keller} B.~W.,
  {Benincasa} S.~M.,  2017, \mn@doi [\mnras] {10.1093/mnras/stx736}, \href
  {https://ui.adsabs.harvard.edu/abs/2017MNRAS.468.4189P} {468, 4189}

\bibitem[\protect\citeauthoryear{{Plummer}}{{Plummer}}{1911}]{1911MNRAS..71..460P}
{Plummer} H.~C.,  1911, \mn@doi [\mnras] {10.1093/mnras/71.5.460}, \href
  {https://ui.adsabs.harvard.edu/abs/1911MNRAS..71..460P} {71, 460}

\bibitem[\protect\citeauthoryear{{Poggio} et~al.,}{{Poggio}
  et~al.}{2018}]{2018MNRAS.481L..21P}
{Poggio} E.,  et~al., 2018, \mn@doi [\mnras] {10.1093/mnrasl/sly148}, \href
  {https://ui.adsabs.harvard.edu/abs/2018MNRAS.481L..21P} {481, L21}

\bibitem[\protect\citeauthoryear{{Purcell}, {Bullock}, {Tollerud}, {Rocha}  \&
  {Chakrabarti}}{{Purcell} et~al.}{2011}]{2011Natur.477..301P}
{Purcell} C.~W.,  {Bullock} J.~S.,  {Tollerud} E.~J.,  {Rocha} M.,
  {Chakrabarti} S.,  2011, \mn@doi [\nat] {10.1038/nature10417}, \href
  {https://ui.adsabs.harvard.edu/abs/2011Natur.477..301P} {477, 301}

\bibitem[\protect\citeauthoryear{{Queiroz} et~al.,}{{Queiroz}
  et~al.}{2023}]{2023A&A...673A.155Q}
{Queiroz} A.~B.~A.,  et~al., 2023, \mn@doi [\aap]
  {10.1051/0004-6361/202245399}, \href
  {https://ui.adsabs.harvard.edu/abs/2023A&A...673A.155Q} {673, A155}

\bibitem[\protect\citeauthoryear{{Ratcliffe} et~al.,}{{Ratcliffe}
  et~al.}{2023}]{2023MNRAS.tmp.1561R}
{Ratcliffe} B.,  et~al., 2023, \mn@doi [\mnras] {10.1093/mnras/stad1573}, \href
  {https://ui.adsabs.harvard.edu/abs/2023MNRAS.525.2208R} {525, 2208}

\bibitem[\protect\citeauthoryear{{Read} \& {Erkal}}{{Read} \&
  {Erkal}}{2019}]{2019MNRAS.487.5799R}
{Read} J.~I.,  {Erkal} D.,  2019, \mn@doi [\mnras] {10.1093/mnras/stz1320},
  \href {https://ui.adsabs.harvard.edu/abs/2019MNRAS.487.5799R} {487, 5799}

\bibitem[\protect\citeauthoryear{{Reid} et~al.,}{{Reid}
  et~al.}{2009}]{2009ApJ...700..137R}
{Reid} M.~J.,  et~al., 2009, \mn@doi [\apj] {10.1088/0004-637X/700/1/137},
  \href {https://ui.adsabs.harvard.edu/abs/2009ApJ...700..137R} {700, 137}

\bibitem[\protect\citeauthoryear{{Renaud}, {Boily}, {Naab}  \&
  {Theis}}{{Renaud} et~al.}{2009}]{2009ApJ...706...67R}
{Renaud} F.,  {Boily} C.~M.,  {Naab} T.,   {Theis} C.,  2009, \mn@doi [\apj]
  {10.1088/0004-637X/706/1/67}, \href
  {https://ui.adsabs.harvard.edu/abs/2009ApJ...706...67R} {706, 67}

\bibitem[\protect\citeauthoryear{{Renaud}, {Bournaud}, {Kraljic}  \&
  {Duc}}{{Renaud} et~al.}{2014}]{2014MNRAS.442L..33R}
{Renaud} F.,  {Bournaud} F.,  {Kraljic} K.,   {Duc} P.~A.,  2014, \mn@doi
  [\mnras] {10.1093/mnrasl/slu050}, \href
  {https://ui.adsabs.harvard.edu/abs/2014MNRAS.442L..33R} {442, L33}

\bibitem[\protect\citeauthoryear{{Renaud}, {Agertz}, {Read}, {Ryde},
  {Andersson}, {Bensby}, {Rey}  \& {Feuillet}}{{Renaud}
  et~al.}{2021}]{2021MNRAS.503.5846R}
{Renaud} F.,  {Agertz} O.,  {Read} J.~I.,  {Ryde} N.,  {Andersson} E.~P.,
  {Bensby} T.,  {Rey} M.~P.,   {Feuillet} D.~K.,  2021, \mn@doi [\mnras]
  {10.1093/mnras/stab250}, \href
  {https://ui.adsabs.harvard.edu/abs/2021MNRAS.503.5846R} {503, 5846}

\bibitem[\protect\citeauthoryear{{Renaud}, {Segovia Otero}  \&
  {Agertz}}{{Renaud} et~al.}{2022}]{2022MNRAS.516.4922R}
{Renaud} F.,  {Segovia Otero} {\'A}.,   {Agertz} O.,  2022, \mn@doi [\mnras]
  {10.1093/mnras/stac2557}, \href
  {https://ui.adsabs.harvard.edu/abs/2022MNRAS.516.4922R} {516, 4922}

\bibitem[\protect\citeauthoryear{{Ruiz-Lara}, {Few}, {Gibson}, {P{\'e}rez},
  {Florido}, {Minchev}  \& {S{\'a}nchez-Bl{\'a}zquez}}{{Ruiz-Lara}
  et~al.}{2016}]{2016A&A...586A.112R}
{Ruiz-Lara} T.,  {Few} C.~G.,  {Gibson} B.~K.,  {P{\'e}rez} I.,  {Florido} E.,
  {Minchev} I.,   {S{\'a}nchez-Bl{\'a}zquez} P.,  2016, \mn@doi [\aap]
  {10.1051/0004-6361/201526470}, \href
  {https://ui.adsabs.harvard.edu/abs/2016A&A...586A.112R} {586, A112}

\bibitem[\protect\citeauthoryear{{Ruiz-Lara}, {Gallart}, {Bernard}  \&
  {Cassisi}}{{Ruiz-Lara} et~al.}{2020}]{2020NatAs...4..965R}
{Ruiz-Lara} T.,  {Gallart} C.,  {Bernard} E.~J.,   {Cassisi} S.,  2020, \mn@doi
  [Nature Astronomy] {10.1038/s41550-020-1097-0}, \href
  {https://ui.adsabs.harvard.edu/abs/2020NatAs...4..965R} {4, 965}

\bibitem[\protect\citeauthoryear{{Ruiz-Lara} et~al.,}{{Ruiz-Lara}
  et~al.}{2021}]{2021MNRAS.501.3962R}
{Ruiz-Lara} T.,  et~al., 2021, \mn@doi [\mnras] {10.1093/mnras/staa3871}, \href
  {https://ui.adsabs.harvard.edu/abs/2021MNRAS.501.3962R} {501, 3962}

\bibitem[\protect\citeauthoryear{{Rupke}, {Kewley}  \& {Barnes}}{{Rupke}
  et~al.}{2010}]{2010ApJ...710L.156R}
{Rupke} D. S.~N.,  {Kewley} L.~J.,   {Barnes} J.~E.,  2010, \mn@doi [\apjl]
  {10.1088/2041-8205/710/2/L156}, \href
  {https://ui.adsabs.harvard.edu/abs/2010ApJ...710L.156R} {710, L156}

\bibitem[\protect\citeauthoryear{{Rusakov}, {Monelli}, {Gallart}, {Fritz},
  {Ruiz-Lara}, {Bernard}  \& {Cassisi}}{{Rusakov}
  et~al.}{2021}]{2021MNRAS.502..642R}
{Rusakov} V.,  {Monelli} M.,  {Gallart} C.,  {Fritz} T.~K.,  {Ruiz-Lara} T.,
  {Bernard} E.~J.,   {Cassisi} S.,  2021, \mn@doi [\mnras]
  {10.1093/mnras/stab006}, \href
  {https://ui.adsabs.harvard.edu/abs/2021MNRAS.502..642R} {502, 642}

\bibitem[\protect\citeauthoryear{{Sch{\"o}nrich} \& {Dehnen}}{{Sch{\"o}nrich}
  \& {Dehnen}}{2018}]{2018MNRAS.478.3809S}
{Sch{\"o}nrich} R.,  {Dehnen} W.,  2018, \mn@doi [\mnras]
  {10.1093/mnras/sty1256}, \href
  {https://ui.adsabs.harvard.edu/abs/2018MNRAS.478.3809S} {478, 3809}

\bibitem[\protect\citeauthoryear{{Siegel} et~al.,}{{Siegel}
  et~al.}{2007}]{2007ApJ...667L..57S}
{Siegel} M.~H.,  et~al., 2007, \mn@doi [\apjl] {10.1086/522003}, \href
  {https://ui.adsabs.harvard.edu/abs/2007ApJ...667L..57S} {667, L57}

\bibitem[\protect\citeauthoryear{{Sohn} et~al.,}{{Sohn}
  et~al.}{2007}]{2007ApJ...663..960S}
{Sohn} S.~T.,  et~al., 2007, \mn@doi [\apj] {10.1086/518302}, \href
  {https://ui.adsabs.harvard.edu/abs/2007ApJ...663..960S} {663, 960}

\bibitem[\protect\citeauthoryear{{Sparre} \& {Springel}}{{Sparre} \&
  {Springel}}{2016}]{2016MNRAS.462.2418S}
{Sparre} M.,  {Springel} V.,  2016, \mn@doi [\mnras] {10.1093/mnras/stw1793},
  \href {https://ui.adsabs.harvard.edu/abs/2016MNRAS.462.2418S} {462, 2418}

\bibitem[\protect\citeauthoryear{{Sparre}, {Whittingham}, {Damle}, {Hani},
  {Richter}, {Ellison}, {Pfrommer}  \& {Vogelsberger}}{{Sparre}
  et~al.}{2022}]{2022MNRAS.509.2720S}
{Sparre} M.,  {Whittingham} J.,  {Damle} M.,  {Hani} M.~H.,  {Richter} P.,
  {Ellison} S.~L.,  {Pfrommer} C.,   {Vogelsberger} M.,  2022, \mn@doi [\mnras]
  {10.1093/mnras/stab3171}, \href
  {https://ui.adsabs.harvard.edu/abs/2022MNRAS.509.2720S} {509, 2720}

\bibitem[\protect\citeauthoryear{{Springel} \& {Hernquist}}{{Springel} \&
  {Hernquist}}{2003}]{2003MNRAS.339..312S}
{Springel} V.,  {Hernquist} L.,  2003, \mn@doi [\mnras]
  {10.1046/j.1365-8711.2003.06207.x}, \href
  {https://ui.adsabs.harvard.edu/abs/2003MNRAS.339..312S} {339, 312}

\bibitem[\protect\citeauthoryear{{Springel}, {Di Matteo}  \&
  {Hernquist}}{{Springel} et~al.}{2005}]{2005MNRAS.361..776S}
{Springel} V.,  {Di Matteo} T.,   {Hernquist} L.,  2005, \mn@doi [\mnras]
  {10.1111/j.1365-2966.2005.09238.x}, \href
  {https://ui.adsabs.harvard.edu/abs/2005MNRAS.361..776S} {361, 776}

\bibitem[\protect\citeauthoryear{{Sysoliatina} \& {Just}}{{Sysoliatina} \&
  {Just}}{2021}]{2021A&A...647A..39S}
{Sysoliatina} K.,  {Just} A.,  2021, \mn@doi [\aap]
  {10.1051/0004-6361/202038840}, \href
  {https://ui.adsabs.harvard.edu/abs/2021A&A...647A..39S} {647, A39}

\bibitem[\protect\citeauthoryear{{Taibi}, {Battaglia}, {Leaman}, {Brooks},
  {Riggs}, {Munshi}, {Revaz}  \& {Jablonka}}{{Taibi}
  et~al.}{2022}]{2022A&A...665A..92T}
{Taibi} S.,  {Battaglia} G.,  {Leaman} R.,  {Brooks} A.,  {Riggs} C.,  {Munshi}
  F.,  {Revaz} Y.,   {Jablonka} P.,  2022, \mn@doi [\aap]
  {10.1051/0004-6361/202243508}, \href
  {https://ui.adsabs.harvard.edu/abs/2022A&A...665A..92T} {665, A92}

\bibitem[\protect\citeauthoryear{{Tepper-Garc{\'\i}a} \&
  {Bland-Hawthorn}}{{Tepper-Garc{\'\i}a} \&
  {Bland-Hawthorn}}{2018}]{2018MNRAS.478.5263T}
{Tepper-Garc{\'\i}a} T.,  {Bland-Hawthorn} J.,  2018, \mn@doi [\mnras]
  {10.1093/mnras/sty1359}, \href
  {https://ui.adsabs.harvard.edu/abs/2018MNRAS.478.5263T} {478, 5263}

\bibitem[\protect\citeauthoryear{{Tollet}, {Cattaneo}, {Mamon}, {Moutard}  \&
  {van den Bosch}}{{Tollet} et~al.}{2017}]{2017MNRAS.471.4170T}
{Tollet} {\'E}.,  {Cattaneo} A.,  {Mamon} G.~A.,  {Moutard} T.,   {van den
  Bosch} F.~C.,  2017, \mn@doi [\mnras] {10.1093/mnras/stx1840}, \href
  {https://ui.adsabs.harvard.edu/abs/2017MNRAS.471.4170T} {471, 4170}

\bibitem[\protect\citeauthoryear{{Torrey}, {Cox}, {Kewley}  \&
  {Hernquist}}{{Torrey} et~al.}{2012}]{2012ApJ...746..108T}
{Torrey} P.,  {Cox} T.~J.,  {Kewley} L.,   {Hernquist} L.,  2012, \mn@doi
  [\apj] {10.1088/0004-637X/746/1/108}, \href
  {https://ui.adsabs.harvard.edu/abs/2012ApJ...746..108T} {746, 108}

\bibitem[\protect\citeauthoryear{{Vall{\'e}e}}{{Vall{\'e}e}}{2005}]{2005AJ....130..569V}
{Vall{\'e}e} J.~P.,  2005, \mn@doi [\aj] {10.1086/431744}, \href
  {https://ui.adsabs.harvard.edu/abs/2005AJ....130..569V} {130, 569}

\bibitem[\protect\citeauthoryear{{Vasiliev}}{{Vasiliev}}{2019}]{2019MNRAS.482.1525V}
{Vasiliev} E.,  2019, \mn@doi [\mnras] {10.1093/mnras/sty2672}, \href
  {https://ui.adsabs.harvard.edu/abs/2019MNRAS.482.1525V} {482, 1525}

\bibitem[\protect\citeauthoryear{{Vasiliev} \& {Belokurov}}{{Vasiliev} \&
  {Belokurov}}{2020}]{2020MNRAS.497.4162V}
{Vasiliev} E.,  {Belokurov} V.,  2020, \mn@doi [\mnras]
  {10.1093/mnras/staa2114}, \href
  {https://ui.adsabs.harvard.edu/abs/2020MNRAS.497.4162V} {497, 4162}

\bibitem[\protect\citeauthoryear{{Vasiliev}, {Belokurov}  \&
  {Erkal}}{{Vasiliev} et~al.}{2021}]{2021MNRAS.501.2279V}
{Vasiliev} E.,  {Belokurov} V.,   {Erkal} D.,  2021, \mn@doi [\mnras]
  {10.1093/mnras/staa3673}, \href
  {https://ui.adsabs.harvard.edu/abs/2021MNRAS.501.2279V} {501, 2279}

\bibitem[\protect\citeauthoryear{{Wegg} \& {Gerhard}}{{Wegg} \&
  {Gerhard}}{2013}]{2013MNRAS.435.1874W}
{Wegg} C.,  {Gerhard} O.,  2013, \mn@doi [\mnras] {10.1093/mnras/stt1376},
  \href {https://ui.adsabs.harvard.edu/abs/2013MNRAS.435.1874W} {435, 1874}

\bibitem[\protect\citeauthoryear{{de Boer}, {Belokurov}  \& {Koposov}}{{de
  Boer} et~al.}{2015}]{2015MNRAS.451.3489D}
{de Boer} T.~J.~L.,  {Belokurov} V.,   {Koposov} S.,  2015, \mn@doi [\mnras]
  {10.1093/mnras/stv946}, \href
  {https://ui.adsabs.harvard.edu/abs/2015MNRAS.451.3489D} {451, 3489}

\bibitem[\protect\citeauthoryear{{de la Vega}, {Quillen}, {Carlin},
  {Chakrabarti}  \& {D'Onghia}}{{de la Vega}
  et~al.}{2015}]{2015MNRAS.454..933D}
{de la Vega} A.,  {Quillen} A.~C.,  {Carlin} J.~L.,  {Chakrabarti} S.,
  {D'Onghia} E.,  2015, \mn@doi [\mnras] {10.1093/mnras/stv2055}, \href
  {https://ui.adsabs.harvard.edu/abs/2015MNRAS.454..933D} {454, 933}

\makeatother
\end{thebibliography}

% Don't change these lines
\bsp	% typesetting comment
\label{lastpage}
\end{document}